\documentclass[journal]{IEEEtran}
\usepackage{amsfonts}
\usepackage{amsmath}
\usepackage{amssymb}
\usepackage{booktabs}
\usepackage{balance}
\usepackage{color}
\usepackage{citesort}
\usepackage{graphicx}
\usepackage{mathrsfs}
\usepackage{mdwmath}
\usepackage{multirow}
\usepackage{setspace}
\usepackage{subfigure}
\normalsize

\ifCLASSINFOpdf
\else
\fi

\newcommand{\CarN}{N}
\newcommand{\CarK}{K}
\newcommand{\CarBlocks}{M}
\newcommand{\CarBlocksBMST}{B}
\newcommand{\basicN}{n}
\newcommand{\basicK}{k}
\newcommand{\Memory}{m}
\newcommand{\diag}{\mbox{diag}}
\newcommand{\Qfun}[1]{{\rm Q}\left(#1\right)}

\ifCLASSOPTIONonecolumn
\newcommand{\figwidth}{0.585\textwidth}
\newcommand{\figwidthMiddle}{0.6\textwidth}
\newcommand{\figwidthLarge}{0.6\textwidth}
\newcommand{\vspaceh}{0cm}
\fi
\ifCLASSOPTIONtwocolumn
\newcommand{\figwidth}{0.48\textwidth}
\newcommand{\figwidthMiddle}{0.475\textwidth}
\newcommand{\figwidthLarge}{0.48\textwidth}
\newcommand{\vspaceh}{0cm}
\fi

\newcommand{\mathbfit}[1]{\mbox{\boldmath$#1$\unboldmath}}

\newtheorem{algorithm}{\textbf{Algorithm}}
\newtheorem{example}{\textbf{Example}}
\makeatletter
    
    \newcommand{\Rmnum}[1]{\expandafter\@slowromancap\romannumeral #1@}
\makeatother

\begin{document}
\title{Performance Analysis of Block Markov Superposition Transmission of Short Codes}
%

\author{Kechao~Huang,~\IEEEmembership{Student~Member,~IEEE,}
        and Xiao~Ma,~\IEEEmembership{Member,~IEEE}
        \thanks{This work was partially supported by the $973$ Program (No. $2012$CB$316100$) and the China NSF (No. 91438101 and No. 61172082).}
        \thanks{The authors are with the Department of Electronics and Communication Engineering, Sun Yat-sen University, Guangzhou, GD 510006, China~(e-mail:~hkech@mail2.sysu.edu.cn; maxiao@mail.sysu.edu.cn).}
}
\markboth{IEEE J. Sel. Areas Commun. (Submitted Paper)}{}%



\maketitle
\IEEEpeerreviewmaketitle

\begin{abstract}
In this paper, we consider the asymptotic and finite-length performance of block Markov superposition transmission~(BMST) of short codes, which can be viewed as a new class of spatially coupled~(SC) codes with the generator matrices of short codes~(referred to as {\em basic codes}) coupled. A modified extrinsic information transfer~(EXIT) chart analysis that takes into account the relation between mutual information~(MI) and bit-error-rate~(BER) is presented to study the convergence behavior of BMST codes. Using the modified EXIT chart analysis, we investigate the impact of various parameters on BMST code performance, thereby providing theoretical guidance for designing and implementing practical BMST codes suitable for sliding window decoding. Then, we present a performance comparison of BMST codes and SC low-density parity-check (SC-LDPC) codes on the basis of equal decoding latency. Also presented is a comparison of computational complexity. Simulation results show that, under the equal decoding latency constraint, BMST codes using the repetition code as the basic code can outperform $(3,6)$-regular SC-LDPC codes in the waterfall region but have a higher computational complexity.
\end{abstract}


\section{Introduction}
Low-density parity-check~(LDPC) block codes~(LDPC-BCs)~\cite{Gallager63}, combined with iterative belief propagation~(BP) decoding, are a class of capacity-approaching codes with decoding complexity that increases only linearly with block length~\cite{Richardson01}. A practical approach to improving the performance of LDPC-BCs is coupling together a series of $L$ disjoint graphs that specify the parity-check matrix of an LDPC-BC into a single coupled chain, thereby producing a spatially coupled LDPC~(SC-LDPC) code. It has been shown in~\cite{Lentmaier10,Kudekar11,Kudekar13,Mitchell14} that SC-LDPC code ensembles exhibit a phenomenon called ``threshold saturation", which allows them to achieve the maximum {\em a posteriori}~(MAP) thresholds of their underlying LDPC-BC ensembles on memoryless binary-input symmetric-output channels under BP decoding, and thus to achieve capacity by increasing the density of the parity-check matrix. Due to their excellent performance, SC-LDPC codes have recently received a great deal of attention in the literature ~(see,~e.g.,~\cite{Pusane11,Lentmaier11,Hassan12_ITW,Andriyanova13,Mitchell13,Mitchell2014_Turbo,Costello14,Olmos14,Huang14} and the references therein).

\begin{figure}[t]
   \centering
   \includegraphics[width=\figwidth]{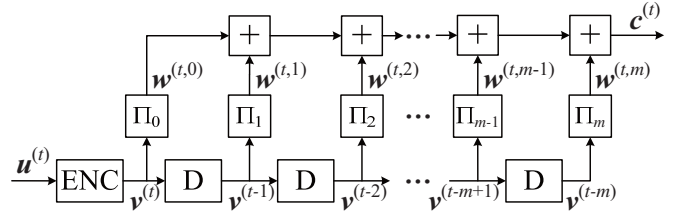}
   \caption{Encoder of a BMST code with encoding memory $\Memory$, where the information sequence $\mathbfit{u}^{(t)}$ at time $t$ is encoded into the sub-codeword $\mathbfit{c}^{(t)}$ for transmission.}
   \label{BMST_encoder}
\end{figure}
The concept of spatial coupling is not limited to LDPC codes. Block Markov superposition transmission~(BMST) of short codes~\cite{Ma15,Liang14c}, for example, is equivalent to spatial coupling of the subgraphs that specify the generator matrices of the short codes. From this perspective, BMST codes are similar to braided block/convolutional codes~\cite{Feltstrom09,Zhang10,Lentmaier14}, staircase codes~\cite{Smith12}, and SC turbo codes~\cite{Moloudi14}. An encoder of a BMST code with encoding memory $\Memory$ is shown in Fig.~\ref{BMST_encoder}, where a BMST code can also be viewed as a serially concatenated code with a structure similar to repeat-accumulate-like codes~\cite{Divsalar98,Pfister03,Abbasfar07}. The outer code is a short code, referred to as the {\em basic code}~(not limited to repetition codes), that introduces redundancy, while the inner code is a rate-one block-oriented feedforward convolutional code~(instead of a bit-oriented accumulator) that introduces memory between transmissions. Hence, BMST codes typically have very simple encoding algorithms. To decode BMST codes, a sliding window decoding algorithm with a tunable decoding delay can be used, as with SC-LDPC codes~\cite{Iyengar12}. The construction of BMST codes is flexible~\cite{Liang14x,Hu14a}, in the sense that it applies to all code rates of interest in the interval (0,1). Further, BMST codes have near-capacity performance~(observed by simulation) in the waterfall region of the bit-error-rate~(BER) cruve and an error floor~(predicted by analysis) that can be controlled by the encoding memory.

On an additive white Gaussian noise channel~(AWGNC), the well-known extrinsic information transfer~(EXIT) chart analysis~\cite{Brink01} can be used to obtain the iterative BP decoding threshold of LDPC-BC ensembles. In~\cite{Liva07}, a novel EXIT chart analysis was used to evaluate the performance of protograph-based LDPC-BC ensembles, and a similar analysis was used to find the thresholds of $q$-ary SC-LDPC codes with sliding window decoding in~\cite{Wei14_IT}. Unlike LDPC codes, the asymptotic BER of BMST codes with window decoding cannot be better than a corresponding genie-aided lower bound~\cite{Ma15}. Thus, conventional EXIT chart analysis cannot be applied directly to BMST codes. In this paper, we propose a modified EXIT chart analysis, that takes into account the relation between mutual information~(MI) and BER, to study the convergence behavior of BMST codes and to predict the performance in the waterfall region of the BER curve. Simulation results confirm that the modified EXIT chart analysis of BMST codes is supported by their finite-length performance behavior. We also investigate the relationship between the basic code structure, the decoding delay, and the decoding performance of BMST codes when the decoding latency is fixed. Finally, we present a computational complexity comparison of BMST codes and SC-LDPC codes on the basis of equal decoding latency.

The rest of the paper is structured as follows. In Section~\ref{SecII}, we give a brief review of BMST codes. In Section~\ref{SecIII}, we discuss the relation between BMST codes and protograph-based SC-LDPC codes. In Section~\ref{SecIV}, we propose a modified EXIT chart analysis of BMST codes. In Section~\ref{SecV}, we investigate the impact of various parameters on BMST code performance. Then, in Section~\ref{SecVI}, we present a performance comparison of BMST codes and SC-LDPC codes on the basis of equal decoding latency. A computational complexity comparison of BMST codes and SC-LDPC codes is also given in Section~\ref{SecVI}. Finally, some concluding remarks are given in Section~\ref{sec:Conclusion}.

\section{Review of BMST Codes}\label{SecII}
\subsection{Encoding of BMST Codes}\label{subsec:encoding}
Consider a BMST code using a rate $R=\basicK/\basicN$ binary basic code $\mathscr{C}[\basicN,\basicK]$ of length $\basicN$ and dimension $\basicK$. Let $\mathbfit{u}=(\mathbfit{u}^{(0)}$, $\mathbfit{u}^{(1)}$, $\cdots$, $\mathbfit{u}^{(L-1)})$ be $L$ blocks of data to be transmitted, where $\mathbfit{u}^{(t)} \in \mathbb{F}_2^k$. Here, $L$ is called the \emph{coupling length}. The encoding algorithm of a BMST code with encoding memory~(\emph{coupling width}) $m$ is described as follows~(see Fig.~\ref{BMST_encoder}), where $\mathbfit{\varPi}_i$ $(0 \leq i \leq \Memory)$ are $m+1$ interleavers of size $n$.

\vspace{0.10cm}
\begin{algorithm}{Encoding of BMST Codes}\label{alg:encoding}
\begin{itemize}
  \item{\bf{Initialization}:} \label{step:encoding_initialize} For $t < 0$, set $\mathbfit{v}^{(t)} = \mathbfit{0} \in \mathbb{F}_2^n$.
  \item{\bf{Loop}:} \label{step:encoding_iteration} For $t = 0$, $1$, $\cdots$, $L-1$,
        \begin{enumerate}
          \item Encode $\mathbfit{u}^{(t)}$ into $\mathbfit{v}^{(t)} \in \mathbb{F}_2^n$ using the encoding algorithm of the basic code $\mathscr{C}$;
          \item For $0\leq i \leq m$, interleave $\mathbfit{v}^{(t-i)}$ using the $i$-th interleaver $\mathbfit{\varPi}_{i}$ into $\mathbfit{w}^{(t,i)}$;
          \item Compute $\mathbfit{c}^{(t)} = \sum_{0\leq i \leq m} \mathbfit{w}^{(t,i)}$, which is taken as the $t$-th block of transmission.
        \end{enumerate}
  \item{\bf{Termination}:} \label{step:encoding_termination}
        For $t = L$, $L+1$, $\cdots$, $L+m-1$, set $ \mathbfit{u}^{(t)} = \mathbfit{0} \in \mathbb{F}_2^k$ and compute $\mathbfit{c}^{(t)}$ following~{\bf Loop}.
  \end{itemize}
\end{algorithm}
\vspace{0.10cm}

\textbf{Remark:} To force the encoder of BMST codes to the zero state at the end of the encoding process, a tail consisting of $m$ blocks of the $k$-dimensional all-zero vector is added. This is different from SC-LDPC code encoders, where the tail is usually non-zero and depends on the encoded information bits~(see Section~IV of~\cite{Pusane08}). As a result, the termination procedure for BMST codes is much simpler than for SC-LDPC codes.

The rate of the BMST code is
\vspace{\vspaceh}
\begin{eqnarray}\label{R_BMST}
    R_{\rm BMST} = \frac{L \basicK}{(L+\Memory)\basicN} = \frac{L}{L+\Memory}R,
\vspace{\vspaceh}
\end{eqnarray}
which is slightly less than the rate $R=\basicK/\basicN$ of the basic code. However, similar to SC-LDPC codes, this rate loss becomes vanishingly small as $L\rightarrow \infty$.

Though any code~(linear or nonlinear) with a fast encoding algorithm and an efficient soft-in soft-out~(SISO) decoding algorithm can be taken as the basic code, we focus in this paper on the use of the $\CarBlocksBMST$-fold Cartesian product of a repetition~(R) code~(denoted by R $[\CarN,1]$) or a single parity-check~(SPC) code~(denoted by SPC $[\CarN,\CarN-1]$) as the basic code, resulting in a BMST-R code~(denoted by BMST-R $[\CarN,1]$) or a BMST-SPC code~(denoted by BMST-SPC $[\CarN,\CarN-1]$), respectively.\footnote{Using codes constructed by time-sharing between the R code and the SPC code as the basic code, one can construct BMST-RSPC codes for a wide range of code rates. For more details, see~\cite{Hu14a}.} Note that the overall code length of the basic code in this case is $\basicN=\CarBlocksBMST \CarN$ and the overall dimension is $\basicK=\CarBlocksBMST$ or $\CarBlocksBMST (\CarN-1)$.


\subsection{Sliding Window Decoding of BMST Codes}\label{subsec:decoding}
BMST codes can be represented by a Forney-style factor graph, also known as a normal graph~\cite{Forney01}, where edges represent variables and vertices~(nodes) represent constraints. All edges connected to a node must satisfy the specific constraint of the node. A full-edge connects to two nodes, while a half-edge connects to only one node. A half-edge is also connected to a special symbol, called a ``dongle", that denotes coupling to other parts of the transmission system~(say, the channel or the information source)~\cite{Forney01}. There are four types of nodes in the normal graph of BMST codes.
\begin{itemize}
  \item \textbf{Node} $\fbox{+}$: All edges~(variables) connected to node $\fbox{+}$ must sum to the all-zero vector. The message updating rule at node $\fbox{+}$ is similar to that of a check node in the factor graph of a binary LDPC code. The only difference is that the messages on the half-edges are obtained from the channel observations.

  \item \textbf{Node} \fbox{$\Pi_{i}$}: The node \fbox{$\Pi_{i}$} represents the $i$-th interleaver, which interleaves or de-interleaves the input messages.

  \item \textbf{Node} $\fbox{=}$: All edges~(variables) connected to node $\fbox{=}$ must take the same (binary) values. The message updating rule at node $\fbox{=}$ is the same as that of a variable node in the factor graph of a binary LDPC code.

  \item \textbf{Node} \fbox{G}: All edges~(variables) connected to node \fbox{G} must satisfy the constraint specified by the basic code $\mathscr{C}$. The message updating rule at node \fbox{G} can be derived accordingly, where the messages on the half-edges are associated with the information source.
\end{itemize}

\begin{figure}[t]
  \centering
  \includegraphics[angle=270, clip, width=\figwidthLarge]{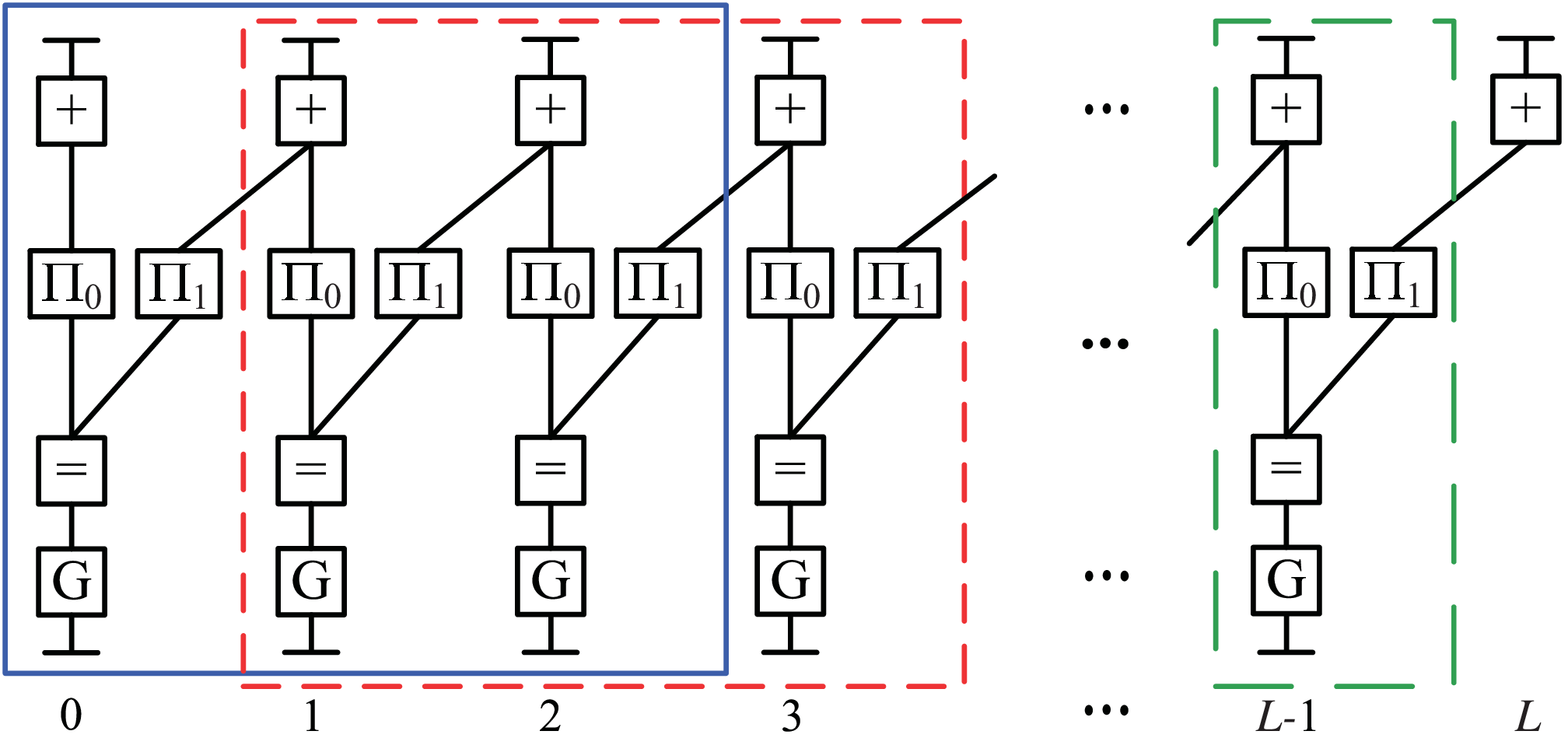}
  \caption{Example of a window decoder with decoding delay $d=2$ operating on the normal graph of a BMST code with $\Memory = 1$ at times $t=0$~(solid blue) and $t=1$~(dotted red). For each window position/time instant, the first~(left-most) decoding layer is called the target layer.}
  \label{fig:decoder}
\end{figure}
The normal graph of a BMST code can be divided into \emph{layers}, where each layer typically consists of a node of type \fbox{G}, a node of type \fbox{=}, $m$ nodes of type \fbox{$\Pi$}, and a node of type \fbox{+} (see Fig.~\ref{fig:decoder}). The result is a high-level normal graph, where each edge represents a sequence of random variables. Looking into the details, we can see that, at each layer, there are $n$ nodes \fbox{$=$} of degree $m+2$, $n$ nodes \fbox{$+$} of degree $m+2$~(including half edges), and $B$ nodes \fbox{${\rm G}_0$} corresponding to the short code~(R $[\CarN,1]$ or SPC $[\CarN,\CarN-1]$ in this paper).

Similar to SC-LDPC codes, an iterative sliding window decoding algorithm with decoding delay $d$ working over a subgraph consisting of $d+1$ consecutive layers can be implemented for BMST codes. An example of a window decoder with decoding delay $d=2$ operating on the normal graph of a BMST code with $\Memory = 1$ is shown in Fig.~\ref{fig:decoder}. For each window position, the forward-backward decoding algorithm is implemented for updating the messages layer-by-layer within the decoding window.\footnote{For more details on the decoding algorithm of BMST codes, we refer the reader to Section~III of~\cite{Ma15}.} Decoding proceeds until a fixed number of iterations has been performed or some given stopping criterion is satisfied, in which case the window shifts to the right by one layer and the symbols corresponding to the layer shifted out of the window are decoded. The first layer in any window is called the {\em target layer}.

\subsection{Genie-Aided Lower Bound on BER}\label{SecIII_Bound}
Let $p_b = f_{\rm BMST}(\gamma_b)$ represent the performance of a BMST code with encoding memory~(coupling width) $\Memory$ and coupling length $L$, where $p_b$ is the BER and $\gamma_b \triangleq E_b/N_0$ represents the received bit signal-to-noise ratio~(SNR) on an AWGNC in dB, and let $p_b = f_{\rm Basic}(\gamma_b)$ represent the performance of the basic code. By assuming a genie-aided decoder, we can obtain a lower bound on the performance of BMST codes given by~(see \cite{Ma15})
\vspace{\vspaceh}
\begin{equation}\label{lowerbound}
    f_{\rm BMST}(\gamma_b) \!\geq\! f_{\rm Basic}\!\left(\! \gamma_b \!+\! 10\!\log_{10}\!\left(\Memory\!+\!1\right) \!-\!10\!\log_{10}\!\left(1 \!+\! \Memory/L\right) \!\right)\!,\!
\end{equation}
where the term $10\log_{10}\left(\Memory+1\right)$ depends on the encoding memory $\Memory$ and the term $10\log_{10}(1+$ $\Memory/L)$ is due to the rate loss. In other words, a maximum coding gain over the basic code of $10\log_{10}(\Memory+1)$ dB in the low BER~(high SNR) region is achieved for large $L$. Intuitively, this bound can be understood by assuming that a codeword in the basic code is transmitted $\Memory+1$ times without interference from other layers.

\subsection{Design of Capacity Approaching BMST Codes}\label{SecIII_Procedure}
Aided by the genie-aided lower bound, we can construct good codes at a target BER with any given code rate of interest by determining as follows the required encoding memory $\Memory$.
\begin{enumerate}
  \item Take a code with the given rate as the basic code. To approach channel capacity, we set the code length $n \geq 10000$;
  \item From the performance curve $f_{\rm Basic}\left(\gamma_b\right)$ of the basic code, find the required $E_b/N_0=\gamma_{\rm target}$ to achieve the target BER;
  \item Find the Shannon limit for the code rate, denoted by $\gamma_{\lim}$;
  \item Determine the encoding memory $m$ by
        \begin{equation}\label{eq:ComputeMemory}
        m = \left\lceil 10^{\frac{\gamma_{\rm target} - \gamma_{\lim}}{10}}-1 \right\rceil,
        \end{equation}
        where $\left\lceil x \right\rceil$ represents the smallest integer greater than or equal to $x$.
\end{enumerate}

The above procedure requires no optimization and hence can be easily implemented given that the performance curve $f_{\rm Basic}\left(\gamma_b\right)$ is available, as is the usual case for short codes.\footnote{The basic code considered in this paper is a Cartesian product of a short code, where each codeword is indeed a cascade of $\CarBlocksBMST$ separate and independent codewords from the short code. Thus, the performance of the basic code can easily be obtained, which is the same as that of the involved short code.} Its effectiveness has been confirmed by construction examples in~\cite{Ma15,Liang14c,Liang14x,Hu14a}. The encoding memories for some BMST codes required to approach the corresponding Shannon limits at given target BERs are shown in Table~\ref{Table1}. As expected, the lower the target BER is, the larger the required encoding memory $\Memory$ is.

\begin{table}
\caption{Encoding memories for BMST codes required to approach the corresponding Shannon limits at given target BERs}\label{Table1}
  \centering
  \begin{tabular}{|c||c|c|c|c|}
  \hline
  \multirow{2}{*}{Encoding memory $\Memory$} &\multicolumn{4}{c|}{Target BER}\\ \cline{2-5}
                                             &$10^{-3}$ &$10^{-4}$ &$10^{-5}$ &$10^{-6}$\\ \hline
  BMST-R $[2,1]$                             &4 &6 &8  &10 \\\hline
  BMST-R $[4,1]$                             &5 &8 &10 &13 \\\hline
  BMST-R $[8,1]$                             &6 &9 &11 &14 \\\hline
  BMST-SPC $[4,3]$                           &2 &3 &4  &5 \\\hline
\end{tabular}
\end{table}

\section{BMST Codes as a Class of SC Codes}\label{SecIII}
In this section, we show that BMST codes can be viewed as a class of SC codes, using an algebraic description as well as a graphical representation, and we compare the structure of BMST codes to SC-LDPC codes.
\subsection{Matrix Representation}
To describe an SC-LDPC code ensemble with coupling width~(\emph{syndrome former memory}) $\Memory$ and coupling length $L$, we start with an $(L+\Memory)(\CarN-\CarK) \times L\CarN$ matrix
\begin{eqnarray}\label{SC-LDPC_ter}
\mathbfit{B}=
    \left[
    \begin{array}{cccc}
        \mathbfit{B}_0     &                 &       & \\
        \mathbfit{B}_1     &\mathbfit{B}_0     &       & \\
        \vdots           &\mathbfit{B}_1     &\ddots & \\
        \mathbfit{B}_{\Memory} &\vdots           &\ddots &\mathbfit{B}_0\\
                         &\mathbfit{B}_{\Memory} &\ddots &\mathbfit{B}_1\\
                         &                 &\ddots &\vdots\\
                         &                 &       &\mathbfit{B}_{\Memory}\\
    \end{array}
    \right],
\end{eqnarray}
where all of the $\Memory+1$ component submatrices $\mathbfit{B}_0,\mathbfit{B}_1,\ldots,\mathbfit{B}_{\Memory}$ have non-negative integer entries and size $(\CarN-\CarK) \times \CarN$. To construct an SC-LDPC code with good performance, we can replace each non-zero entry $b \neq 0$ in $\mathbfit{B}$ with a sum of $b$ nonoverlapping randomly selected $M \times M$ permutation matrices and each zero entry in $\mathbfit{B}$ with the $M \times M$ all-zero matrix, where $b$ is typically a small integer and $M$ is typically a large integer. The resulting SC-LDPC parity-check matrix $\mathbfit{H}_{\rm SC}$ of size $(L+\Memory)(\CarN\!-\!\CarK)M\! \times\! L\CarN M$ is given by
\vspace{\vspaceh}
\ifCLASSOPTIONonecolumn
    \begin{align}\label{H_SC}
\mathbfit{H}_{\rm SC}=
    \begin{bmatrix}
        \mathbfit{H}_0(0)     &                    &       & \\
        \mathbfit{H}_1(1)     &\mathbfit{H}_0(1)     &       & \\
        \vdots              &\mathbfit{H}_{1}(2)   &\ddots & \\
        \mathbfit{H}_{\Memory}(\Memory)   &\vdots              &\ddots &\mathbfit{H}_0(L-1)\\
                            &\mathbfit{H}_{\Memory}(\Memory+1) &\ddots &\mathbfit{H}_1(L)\\
                            &                    &\ddots &\vdots\\
                            &                    &       &\mathbfit{H}_{\Memory}(L+\Memory-1)\\
    \end{bmatrix},
\end{align}
\fi
\ifCLASSOPTIONtwocolumn
    \begin{align}\label{H_SC}
\mathbfit{H}_{\rm SC}= ~~~~~~~~~~~~~~~~~~~~~~~~~~~~~~~~~~~~~~~~~~~~~~~~~~~~ \nonumber \\
    \begin{bmatrix}
        \mathbfit{H}_0(0)     &                    &       & \\
        \mathbfit{H}_1(1)     &\mathbfit{H}_0(1)     &       & \\
        \vdots              &\mathbfit{H}_{1}(2)   &\ddots & \\
        \mathbfit{H}_{\Memory}(\Memory)   &\vdots              &\ddots &\mathbfit{H}_0(L-1)\\
                            &\mathbfit{H}_{\Memory}(\Memory+1) &\ddots &\mathbfit{H}_1(L)\\
                            &                    &\ddots &\vdots\\
                            &                    &       &\mathbfit{H}_{\Memory}(L+\Memory-1)\\
    \end{bmatrix},
\end{align}
\fi
where the blank spaces in $\mathbfit{H}_{\rm SC}$ correspond to zeros and the submatrices $\mathbfit{H}_i(t+i)$ have size $(\CarN-\CarK)M \times \CarN M$, for $0\leq i \leq m$ and $0\leq t \leq L-1$.

In contrast to SC-LDPC codes, it is convenient to describe BMST codes using generator matrices. Let $\mathbfit{G}_0$ be the generator matrix of a short code with dimension $\CarK$ and length $\CarN$. To describe a BMST code ensemble with coupling width~(\emph{encoding memory}) $\Memory$ and coupling length $L$, we start with the $L \times (L+\Memory)$ matrix
\vspace{\vspaceh}
\begin{align}\label{G_basic}
\mathbfit{A} =
        \begin{bmatrix}
           1 &1 &\cdots &1 & &  \\
             &1 &1      &\cdots  &1 &  \\
             &  &\ddots &\ddots  &\ddots &\ddots &\\
             &  &       &1       &1      &\cdots &1 \\
             &  &       &        &1      &1 &\cdots &1 \\
        \end{bmatrix},
\end{align}
which has constant weight $\Memory+1$ in each row. This matrix \mathbfit{A} plays a similar role for constructing BMST codes as the matrix \mathbfit{B} does for constructing SC-LDPC codes. To construct a BMST code with good performance, each nonzero entry $A_{j,j+i}$~($0 \leq j \leq L-1$ and $0 \leq i \leq \Memory$) in $\mathbfit{A}$ is replaced with a matrix $\mathbfit{G}\mathbfit{\varPi}_{i}$, where
\begin{equation}\label{G_basic_code}
\mathbfit{G} =
  \diag\{\underbrace{\mathbfit{G}_0, \cdots, \mathbfit{G}_0}_\CarBlocksBMST \}
\end{equation}
is the generator matrix of the $\CarBlocksBMST$-fold Cartesian product of the short code, the $\mathbfit{\varPi}_i$ ($0 \leq i \leq \Memory$) are $\Memory+1$ randomly selected $\CarN\CarBlocksBMST \times \CarN\CarBlocksBMST$ permutation matrices, and the Cartesian product order $\CarBlocksBMST$ is typically large.
The resulting BMST code has length $(L+\Memory)\CarN\CarBlocksBMST$ and dimension $L\CarK\CarBlocksBMST$, and the generator matrix $\mathbfit{G}_{\rm BMST}$ is given by
\vspace{\vspaceh}
\ifCLASSOPTIONonecolumn
\begin{align}\label{G_BMST}
\mathbfit{G}_{\rm BMST} =
    \begin{bmatrix}
           \mathbfit{G}\mathbfit{\varPi}_{0}       & \mathbfit{G}\mathbfit{\varPi}_{1} & \cdots &
           \mathbfit{G}\mathbfit{\varPi}_{\Memory} &                    &  \\
                              & \mathbfit{G}\mathbfit{\varPi}_{0}       & \mathbfit{G}\mathbfit{\varPi}_{1} &
           \cdots             & \mathbfit{G}\mathbfit{\varPi}_{\Memory} &  \\
                              &                     & \ddots &
           \ddots             & \ddots              & \ddots  &\\
                              &                     &        &
           \mathbfit{G}\mathbfit{\varPi}_{0}       & \mathbfit{G}\mathbfit{\varPi}_{1}  & \cdots &
           \mathbfit{G}\mathbfit{\varPi}_{\Memory} \\
        \end{bmatrix}.
\end{align}
\fi
\ifCLASSOPTIONtwocolumn
\begin{align}\label{G_BMST}
\mathbfit{G}_{\rm BMST} = ~~~~~~~~~~~~~~~~~~~~~~~~~~~~~~~~~~~~~~~~~~~~~~~~~~~~~~~~~ \nonumber \\
    \begin{bmatrix}
           \mathbfit{G}\mathbfit{\varPi}_{0}       & \mathbfit{G}\mathbfit{\varPi}_{1} & \cdots &
           \mathbfit{G}\mathbfit{\varPi}_{\Memory} &                    &  \\
                              & \mathbfit{G}\mathbfit{\varPi}_{0}       & \mathbfit{G}\mathbfit{\varPi}_{1} &
           \cdots             & \mathbfit{G}\mathbfit{\varPi}_{\Memory} &  \\
                              &                     & \ddots &
           \ddots             & \ddots              & \ddots  &\\
                              &                     &        &
           \mathbfit{G}\mathbfit{\varPi}_{0}       & \mathbfit{G}\mathbfit{\varPi}_{1}  & \cdots &
           \mathbfit{G}\mathbfit{\varPi}_{\Memory} \\
        \end{bmatrix}.
\end{align}
\fi
\subsection{Graphical Representation}
SC-LDPC code ensembles are often described in terms of a protograph, where an \emph{edge-spreading} operation is applied to couple a sequence of disjoint block code protographs into a single chain~\cite{Mitchell14}. Usually, no extra edges are introduced during the coupling process. In this paper, we describe the coupling process from a new perspective, where extra edges are allowed to be added. We believe that this new treatment is more general. For example, SC turbo codes~\cite{Moloudi14} are obtained by adding edges to connect each turbo code graph to one or more nearby graphs in the chain. Based on this perspective, we can redescribe SC-LDPC codes as follows.

\begin{figure}[t]
    \center
    \includegraphics[angle=270, clip, width=\figwidthMiddle]{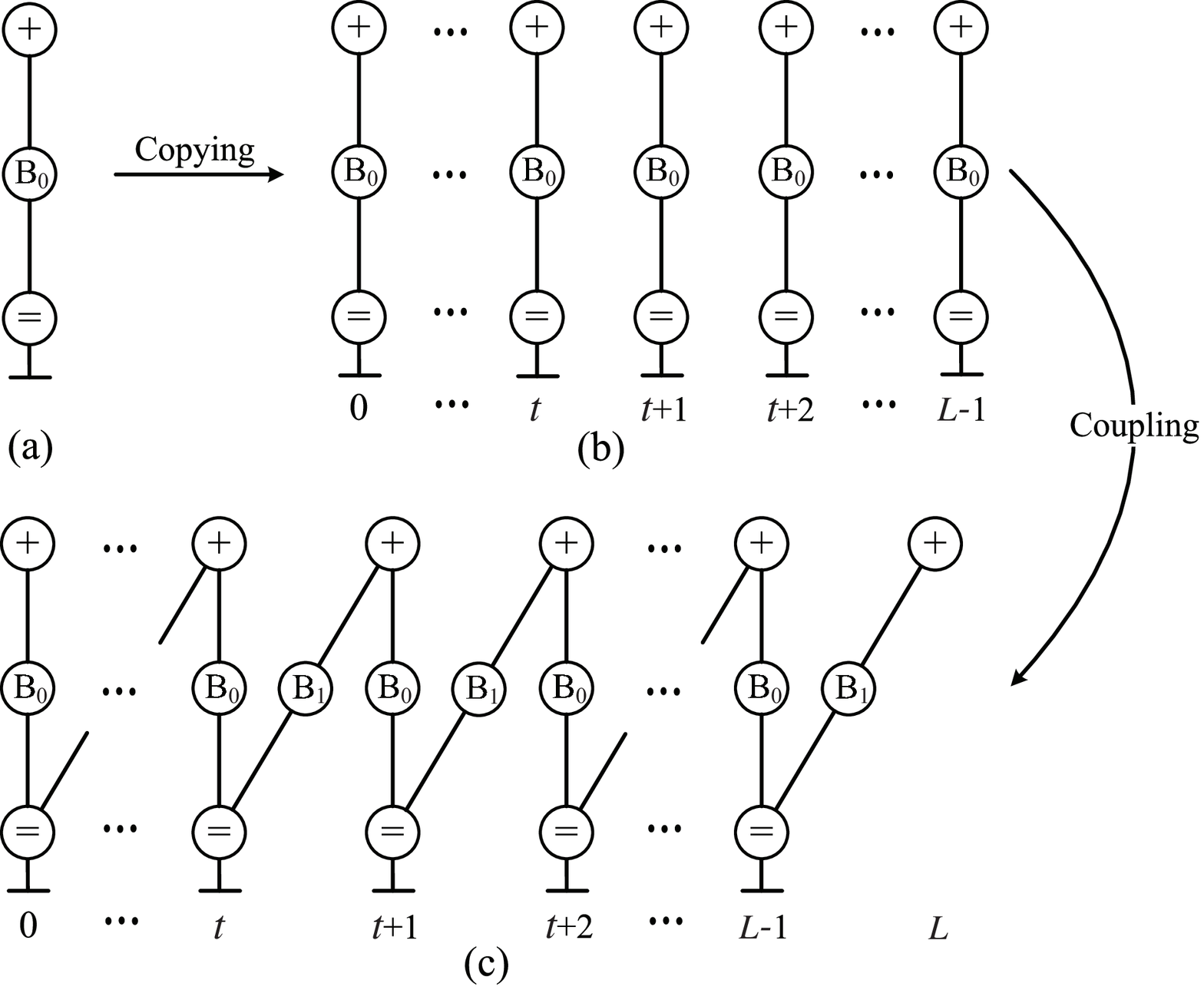}%
    \caption{(a) A protograph corresponding to the submatrix $\mathbfit{B}_{0}$, (b) $L$ uncoupled protographs, each of which corresponds to the submatrix $\mathbfit{B}_{0}$, and (c) a protograph corresponding to an SC-LDPC code ensemble with coupling length $L$ and coupling width $\Memory=1$.}
    \label{Normal_SCLDPC}
\end{figure}
\begin{figure}[t]
    \center
    \includegraphics[clip, width=\figwidthMiddle]{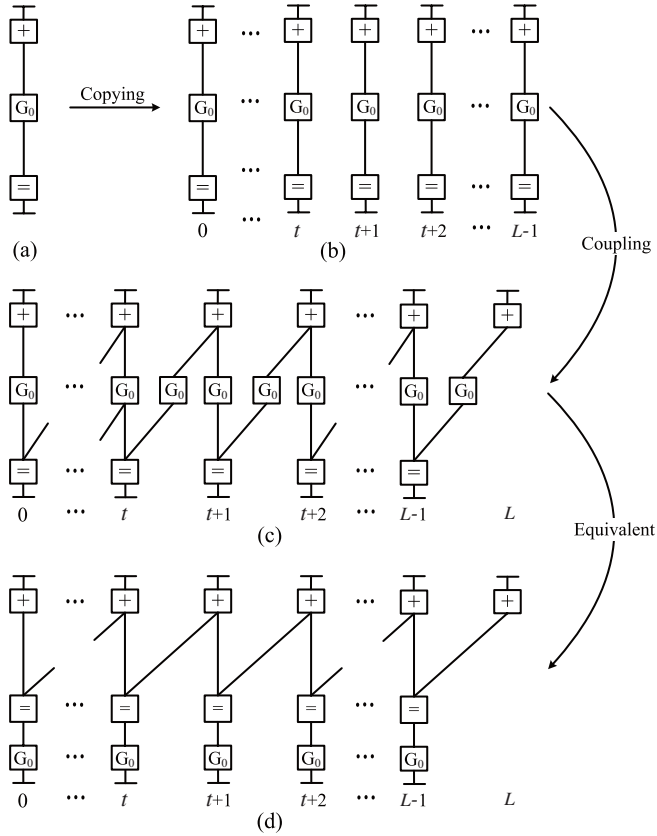}%
    \caption{(a) A protograph representing the basic code with generator matrix $\mathbfit{G}_{0}$, (b) $L$ uncoupled basic protographs, each of which corresponds to the generator matrix $\mathbfit{G}_{0}$, (c) a protograph corresponding to a BMST code ensemble with coupling length $L$ and coupling width $\Memory=1$, and (d) an equivalent protograph corresponding to the same BMST code ensemble with coupling length $L$ and coupling width $\Memory=1$.}
    \label{Normal_BMST}
\end{figure}
We start with a protograph for the submatrix $\mathbfit{B}_{0}=[B_{i,j}]$, which has $\CarN$ variable nodes and $\CarN-\CarK$ check nodes, where the $i$-th check node is connected to the $j$-th variable node by $B_{i,j}$ edges. A short-hand protograph corresponding to $\mathbfit{B}_{0}$ is shown in~Fig.~\ref{Normal_SCLDPC}(a), where the node \textcircled{=} represents $\CarN$ variable nodes, the node \textcircled{+} represents $\CarN-\CarK$ check nodes, and the edge {\large\textcircled{\footnotesize ${\rm B}_0$}} represents a collection of $\sum B_{i,j}$ edges. To distinguish, the edge {\large\textcircled{\footnotesize ${\rm B}_0$}} is referred to as a \emph{super-edge} of type $\mathbfit{B}_{0}$, while the conventional edge in the full protograph is referred to as a \emph{simple edge}. The short-hand protograph is then replicated $L$ times, as shown in~Fig.~\ref{Normal_SCLDPC}(b), meaning that the sequence of transmitted codewords satisfy independently the constraint $\mathbfit{B}_{0}$. The $L$ disjoint graphs are then coupled by adding a \emph{super-edge} of type $\mathbfit{B}_{i}$ to bridge the variable node \textcircled{=} at time $t$ and the check node \textcircled{+} at time $t+i$, for $0 \leq t \leq L-1$ and $1\leq i \leq m$, resulting in a single coupled chain corresponding to an SC-LDPC code ensemble with coupling length $L$ and coupling memory $\Memory$. An example of an SC-LDPC code ensemble with coupling memory $\Memory=1$ is shown in~Fig.~\ref{Normal_SCLDPC}(c). When lifting, each \emph{simple edge}~(not \emph{super-edge}) is replaced by a bundle of $M$ edges~(permutation within the bundle is assumed), resulting in an SC-LDPC code with length $L \CarN \CarBlocks$.

Similarly, BMST codes start with a protograph for the generator matrix $\mathbfit{G}_{0}=[G_{i,j}]$, which has $\CarK$ \fbox{=} nodes and $\CarN$ \fbox{+} nodes, where the $i$-th \fbox{=} node is connected to the $j$-th \fbox{+} node if and only if $G_{i,j}=1$. A short-hand protograph corresponding to $\mathbfit{G}_{0}$ is shown in~Fig.~\ref{Normal_BMST}(a), where \fbox{${\rm G}_0$} represents a \emph{super-edge} of type $\mathbfit{G}_{0}$. The protograph is then replicated $L$ times, as shown in~Fig.~\ref{Normal_BMST}(b), which can be considered as transmitting a sequence of codewords from the basic code corresponding to the generator matrix $\mathbfit{G}_{0}$ independently at time instants $t = 0$, $1$, $\cdots$, $L-1$. The $L$ disjoint graphs are coupled by adding a \emph{super-edge} of type $\mathbfit{G}_{0}$ to bridge the \fbox{$=$} node at time $t$ and the \fbox{$+$} node at time $t+i$, for $1\leq i \leq m$, resulting in a single coupled chain corresponding to a BMST code ensemble with coupling length $L$ and coupling memory $\Memory$. An example of a BMST code ensemble with coupling memory $\Memory=1$ is shown in~Fig.~\ref{Normal_BMST}(c), whose equivalent form is shown in Fig.~\ref{Normal_BMST}(d). When lifting, the \emph{super-edge} of type $\mathbfit{G}_{0}$ bridging the \fbox{$=$} node at time $t$ and the \fbox{$+$} node at time $t+i$, for $0 \leq t \leq L-1$ and $0 \leq i \leq \Memory$, is replaced by a \emph{super-edge} of type $\mathbfit{G}\mathbfit{\varPi}_{i}$, resulting in a BMST code with length $(L+\Memory) \CarN \CarBlocksBMST$.

\subsection{Similarities and Differences}
From the previous two subsections, we see that both SC-LDPC codes and BMST codes can be derived from a small matrix by replacing the entries with properly-defined submatrices. We also see that the generator matrix $\mathbfit{G}_{\rm BMST}$ of BMST codes is similar in form to the parity-check matrix $\mathbfit{H}_{\rm SC}$ of SC-LDPC codes. SC-LDPC codes introduce memory by spatially coupling the basic parity-check matrices $\mathbfit{B}_{0}$, while BMST codes introduce memory by spatially coupling the basic generator matrices $\mathbfit{G}_{0}$. Further, we see from Fig.~\ref{Normal_SCLDPC} and Fig.~\ref{Normal_BMST} that during the construction of both SC-LDPC codes and BMST codes, the memory is introduced by coupling the disjoint graphs together in a single chain, which is the fundamental idea of spatial coupling. Thus, BMST codes can be viewed as a class of SC codes.

\section{EXIT Chart Analysis of BMST Codes}\label{SecIV}
Given the basic code with generator matrix $\mathbfit{G}_0$, we can construct a sequence of BMST codes by choosing the Cartesian product order $\CarBlocksBMST=1,2,\cdots$. Now assume that the interleavers are chosen uniformly and at random for each transmission. Then we have a sequence of code ensembles. The aim of EXIT chart analysis is to predicte the performance behavior of the BMST codes as $\CarBlocksBMST\rightarrow \infty$. In this section, we first discuss the issue that prevents the use of conventional EXIT chart analysis for BMST codes, and then we provide a modified EXIT chart analysis to study the convergence behavior of BMST codes with window decoding.

We consider binary phase-shift keying~(BPSK) modulation over the binary-input AWGNC. To describe density evolution, it is convenient to assume that the all-zero codeword is transmitted and to represent the messages as log-likelihood ratios~(LLRs). The threshold of protograph-based LDPC codes can be obtained based on a protograph-based EXIT chart analysis~\cite{Liva07,Wei14_IT} by determining the minimum value of the SNR $E_b/N_0$ such that the MI between the \emph{a posteriori} message at a variable node and an associated codeword bit~(referred to as the \emph{a posteriori} MI for short) goes to 1 as the number of iterations increases, i.e., the BER at the variable nodes tends to zero as the number of iterations tends to infinity. At a first glance, a similar iterative sliding window decoding EXIT chart analysis algorithm can be implemented over the normal graph~(see Fig.~\ref{Normal_BMST}(d)) of the BMST code ensemble to study the convergence behavior of BMST codes. However, as shown in~(\ref{lowerbound}), the high SNR performance of BMST codes with window decoding cannot be better than the corresponding genie-aided lower bound, which means that the \emph{a posteriori} MI of BMST codes cannot reach 1 as the number of iterations tends to infinity. Thus, the conventional EXIT chart analysis cannot be applied directly to BMST codes. Fortunately, this can be amended by taking into account the relation between MI and BER~\cite{Brink01}. Specifically, we need the convergence check at node \fbox{${\rm G}_0$}, as described below in Algorithm~\ref{alg:convergence}. For convenience, the MI between the \emph{a priori} input and the corresponding codeword bit is referred to as the \emph{a priori MI}, the MI between the \emph{extrinsic} output and the corresponding codeword bit is referred to as the \emph{extrinsic} MI, and the MI between the channel observation and the corresponding codeword bit is referred to as the \emph{channel} MI.


\vspace{0.1cm}
\begin{algorithm}{Convergence Check at Node \fbox{${\rm G}_0$}}\label{alg:convergence}
\begin{itemize}
  \item Let $I_A$ denote the \emph{a priori} MI and $I_E$ denote the \emph{extrinsic} MI. Then the \emph{a posteriori} MI $I_{\rm AP}$ is given by
    \begin{equation}\label{I_APP}
        I_{\rm AP} = J(\sqrt{[J^{-1}(I_A)]^2+[J^{-1}(I_E)]^2}),
    \end{equation}
    where the $J(\cdot)$ and $J^{-1}(\cdot)$ functions are given in~\cite{Brink04}, $I_A$ is the \emph{a priori} MI, and $I_E$ is the \emph{extrinsic} MI. As shown in Section~III-C of~\cite{Brink01}, supposing that the \emph{a posteriori} MI is Gaussian, an estimate of the BER $p_{est}$ is then given by
    \begin{equation}\label{I_APP_BER}
        p_{est} = \Qfun{J^{-1}(1-I_{\rm AP})/2},
    \end{equation}
    where
    \begin{equation}\label{eq:Qfunction}
        \Qfun{x} = \frac{1}{\sqrt{2\pi}}\int_{x}^{\infty}\exp\left(-\frac{t^2}{2}\right)dt.
    \end{equation}

  \item If the estimated BER $p_{est}$ is less than some preselected target BER, a local decoding success is declared; otherwise, a local decoding failure is declared.
\end{itemize}
\end{algorithm}

For a fixed SNR ${E_b}/{N_0}$, the channel bit LLR corresponding to the binary-input AWGNC is Gaussian with variance~\cite{Brink01}
\begin{equation}\label{sigma_ch}
    \sigma^2_{\rm ch}=8 R_{\rm BMST}\frac{E_b}{N_0},
\end{equation}
where $R_{\rm BMST}$ is the rate of the BMST codes. The channel MI is then given by
\begin{equation}\label{J_ch}
    I_{ch}=J\left(\sigma_{\rm ch}\right)=J\left(\sqrt{8 R_{\rm BMST}\frac{E_b}{N_0}}\right).
\end{equation}
The modified EXIT chart analysis algorithm of BMST codes, similar to the protograph-based EXIT chart analysis algorithm of SC-LDPC codes~\cite{Wei14_IT}, can now be described as follows.

\vspace{0.1cm}
\begin{algorithm}{EXIT Chart Analysis of BMST Codes with Window Decoding}\label{alg:decoding}
\begin{itemize}
  \item {\bf{Initialization}:} All messages over those half-edges (connected to the channel) at nodes $\fbox{+}$ are initialized as $I_{ch}$ according to~(\ref{J_ch}), all messages over those half-edges (connected to the information source) at nodes \fbox{${\rm G}_0$} are initialized as 0, and all messages over the remaining (inter-connected) full-edges are initialized as 0. Set a maximum number of iterations $I_{\max}$.

  \item {\bf{Sliding window decoding}:} For each window position, the $d+1$ decoding layers perform MI message processing/passing layer-by-layer according to the schedule
    \begin{equation*}
        \fbox{+} \rightarrow \fbox{=} \rightarrow
        \fbox{${\rm G}_0$} \rightarrow \fbox{=} \rightarrow \fbox{+}.
    \end{equation*}
    After a fixed number of iterations $I_{\max}$, perform a convergence check at node \fbox{${\rm G}_0$} using Algorithm~\ref{alg:convergence}. If a local decoding failure is declared, then window decoding terminates; otherwise, a local decoding success is declared, the window position is shifted, and decoding continues. A complete decoding success for a specific channel parameter ${E_b}/{N_0}$ and target BER is declared if and only if all target layers declare decoding successes.
\end{itemize}
\end{algorithm}

Now we can denote the iterative decoding threshold ${({E_b}/{N_0})}^*$ of BMST code ensembles for a preselected target BER as the minimum value of the channel parameter ${E_b}/{N_0}$ which allows the decoder of Algorithm~\ref{alg:decoding} to output a decoding success, in the limit of large code lengths~(i.e., $\CarBlocksBMST\rightarrow \infty$).

\section{Impact of Parameters on BMST Codes}\label{SecV}
In this section we study the impact of various parameters~(coupling width $\Memory$, Cartesian product order $\CarBlocksBMST$, and decoding delay $d$) on BMST codes. Three regimes are considered: (1)~fixed $\Memory$ and $\CarBlocksBMST$, increasing $d$, (2)~fixed $\Memory$ and $d$, increasing $\CarBlocksBMST$, and (3)~fixed $\CarBlocksBMST$, increasing $\Memory$~(and hence $d$).

All simulations are performed assuming BPSK modulation and an AWGNC. In the computation of the asymptotic window decoding thresholds of BMST codes, we set a maximum number of iterations $I_{\max}=1000$. We will refer to the iterative decoding threshold ${({E_b}/{N_0})}^*$ simply as ${E_b}/{N_0}$ when it does not lead to ambiguity. In the simulations of finite-length performance, $\Memory+1$ random interleavers~(randomly generated but fixed) of size $n=\CarN\CarBlocksBMST$ are used for encoding. The iterative sliding window decoding algorithm~\cite[Algorithm 3]{Ma15} for BMST codes is performed using a layer-by-layer updating schedule with a maximum iteration number of 18, and the entropy stopping criterion~\cite{Ma04,Ma15} with a preselected threshold of $10^{-6}$ is employed.

\ifCLASSOPTIONonecolumn
\begin{figure*}[t]
\fi
\ifCLASSOPTIONtwocolumn
\begin{figure}[t]
\fi
\centering
\subfigure[~]{
\label{RC21DiffDelay}
\includegraphics[clip, width=\figwidth]{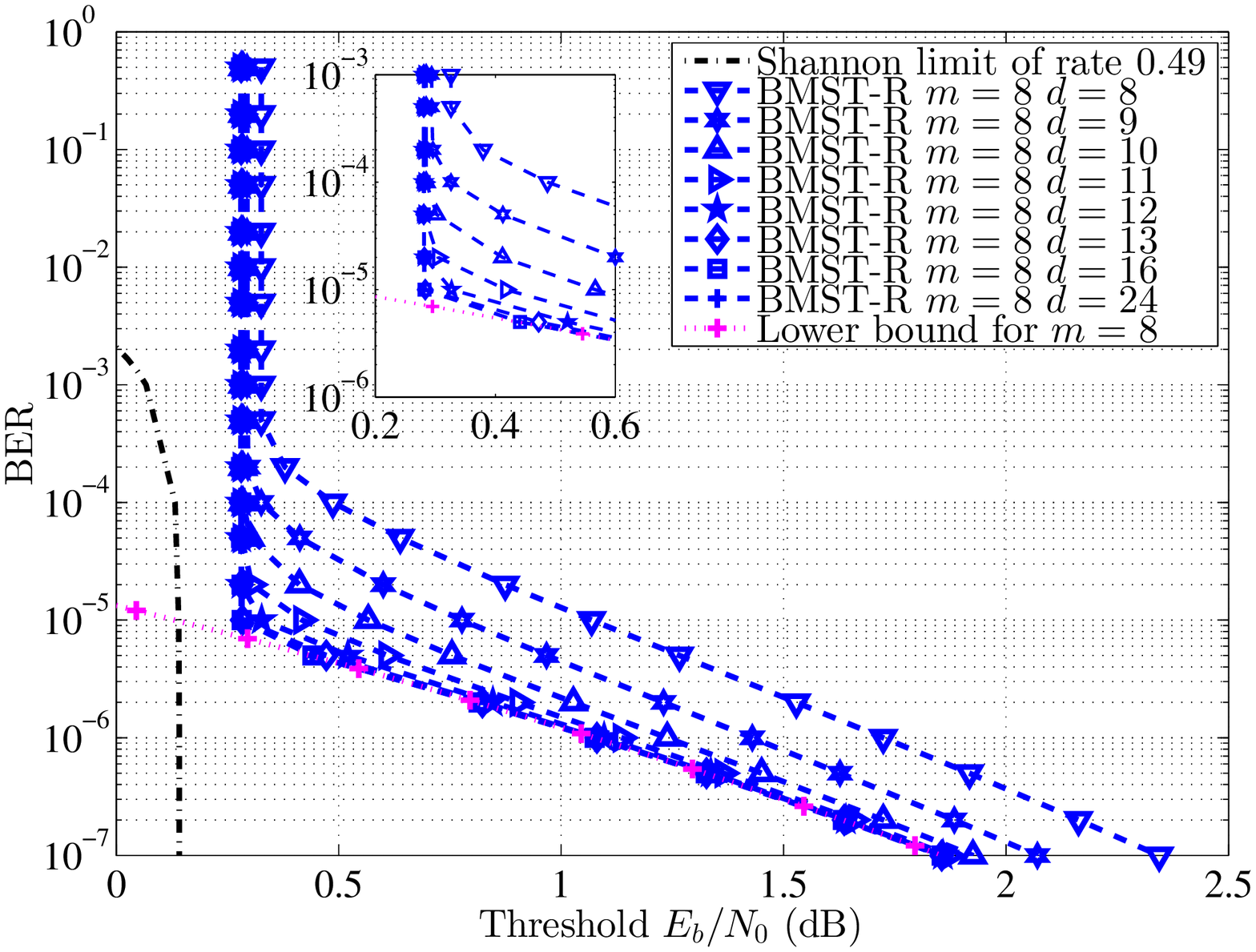}}
\subfigure[~]{
\label{SPC43DiffDelay}
\includegraphics[clip, width=\figwidth]{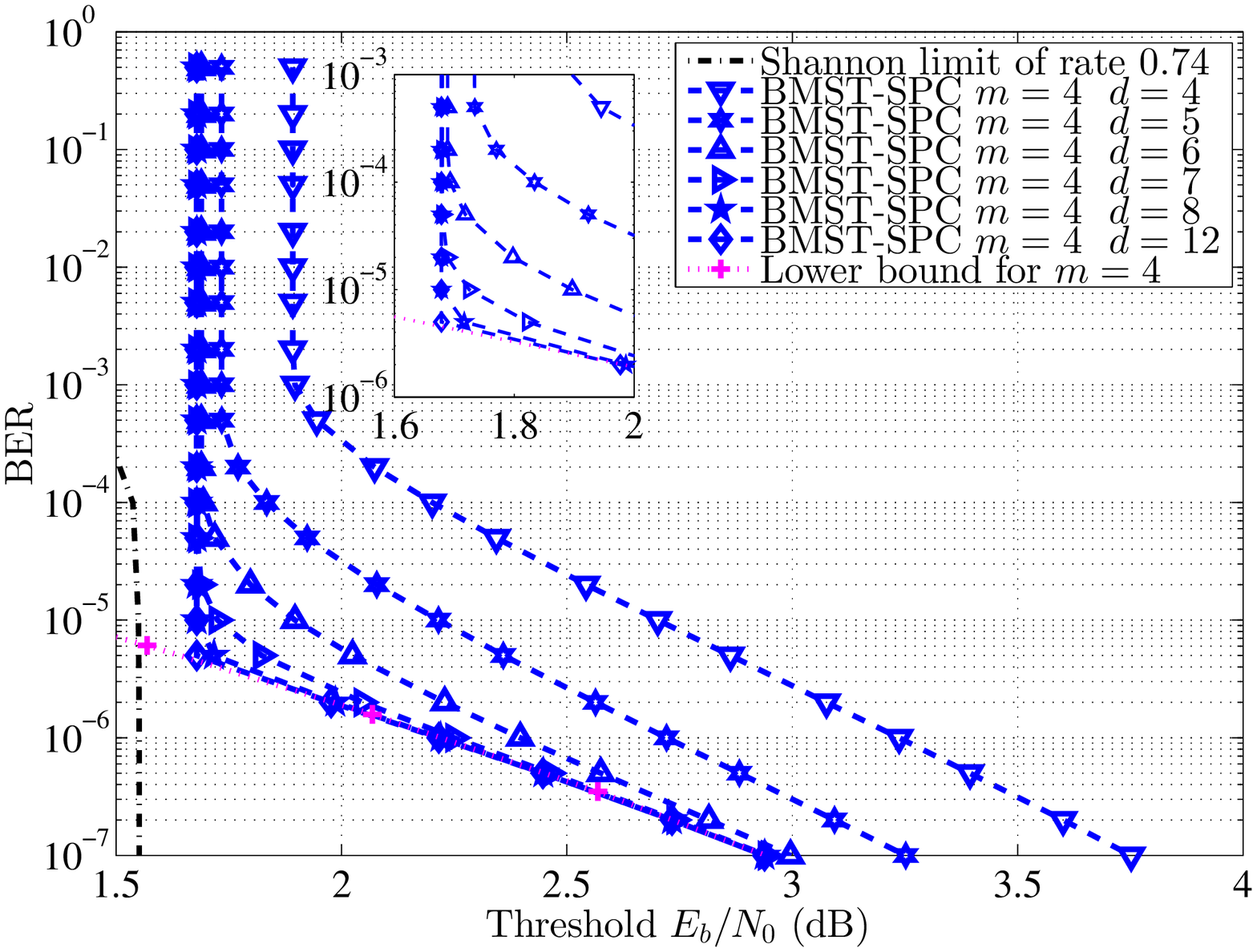}}
\caption{Window decoding thresholds in terms of ${E_b}/{N_0}$ (dB) with different target BERs and different decoding delays for (a)~a rate $R_{\rm BMST}=0.49$ BMST-R $[2,1]$ code ensemble with $\Memory=8$ and $L=392$, and (b)~a rate $R_{\rm BMST}=0.74$ BMST-SPC $[4,3]$ code ensemble with $\Memory=4$ and $L=296$.}
\label{Threshold_RC21_SPC43_DiffDelay}
\ifCLASSOPTIONonecolumn
\end{figure*}
\fi
\ifCLASSOPTIONtwocolumn
\end{figure}
\fi

\begin{figure}[t]
\centering
\includegraphics[clip, width=\figwidth]{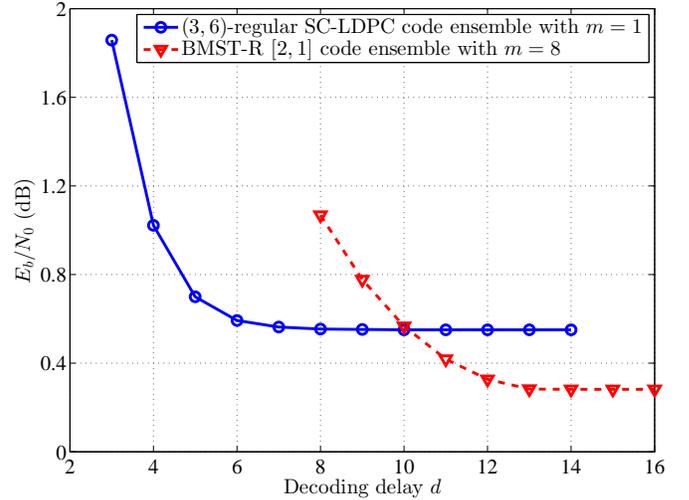}
\caption{Window decoding thresholds for a $(3,6)$-regular SC-LDPC code ensemble with $m=1$ and a BMST-R $[2,1]$ code ensemble with $\Memory=8$ as a function of decoding delay $d$. The preselected target BER is $10^{-5}$. The component submatrices for the SC-LDPC code ensemble are $\mathbfit{B}_0=[2~1]$ and $\mathbfit{B}_1=[1~2]$.}
\label{Threshold_BMST_SCLDPC_Function_Delay}
\end{figure}

\subsection{Fixed $\Memory$ and $\CarBlocksBMST$, Increasing $d$}

\begin{example}[Asymptotic Performance]
Consider a $R_{\rm BMST}=0.49$ BMST-R $[2,1]$ code ensemble with $\Memory=8$ and $L=392$. We calculate its window decoding thresholds with different preselected target BERs and different decoding delays. The calculated thresholds in terms of the SNR ${E_b}/{N_0}$ versus the preselected target BERs together with the lower bound are shown in Fig.~\ref{Threshold_RC21_SPC43_DiffDelay}(a), where we observe that
\begin{enumerate}
  \item In the waterfall region~(above a critical BER), the thresholds remain almost constant. However, once the critical BER is reached, the thresholds increase as the target BER decreases.

  \item For a small decoding delay~(say $d=\Memory$), the thresholds do not achieve the lower bound even in the high SNR region.
  \item For a larger decoding delay~(roughly $d=2\Memory\thicksim3\Memory$), the thresholds correspond to the lower bound in the high SNR region, suggesting that the window decoding algorithm with decoding delay $d\geq2\Memory\thicksim3\Memory$ is near optimal for BMST codes.
  \item The error floor region threshold improves as the decoding delay $d$ increases, but it does not improve much further beyond a certain decoding delay~(roughly $d=2\Memory\thicksim3\Memory$).
\end{enumerate}

Similar behavior has also been observed for BMST-SPC code ensembles, as shown in Fig.~\ref{Threshold_RC21_SPC43_DiffDelay}(b), where the thresholds of a rate $R_{\rm BMST}=0.74$ BMST-SPC $[4,3]$ code ensemble constructed with $\Memory=4$ and $L=296$ and decoded with different decoding delays $d$ are depicted.

The window decoding thresholds, corresponding to a preselected target BER\footnote{We choose a BER of $10^{-5}$ for comparison because it represents a target BER commonly used in many practical applications.} of $10^{-5}$, for the $(3,6)$-regular SC-LDPC code ensemble with $m=1$ and the BMST-R $[2,1]$ code ensemble with $\Memory=8$ as a function of decoding delay $d$ is shown in Fig.~\ref{Threshold_BMST_SCLDPC_Function_Delay}. We see that, similar to the SC-LDPC code ensemble, the threshold of the BMST code ensemble improves as the decoding delay $d$ increases and it becomes better than that of the SC-LDPC code ensemble beyond a certain decoding delay~(roughly $d=10$).
\end{example}

\ifCLASSOPTIONonecolumn
\begin{figure*}[t]
\fi
\ifCLASSOPTIONtwocolumn
\begin{figure}[t]
\fi
\centering
\subfigure[~]{
\label{RC21}
\includegraphics[clip, width=\figwidth]{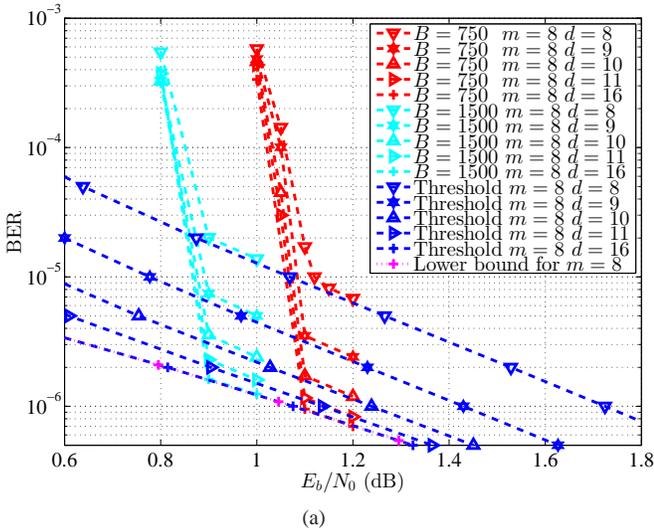}}
\subfigure[~]{
\label{SPC43}
\includegraphics[clip, width=\figwidth]{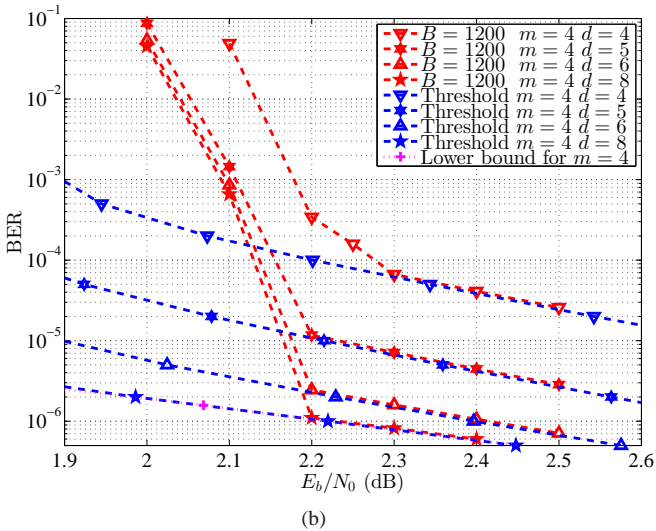}}
\caption{Simulated decoding performance of BMST codes decoded with different decoding delays $d$. The corresponding window decoding thresholds and the lower bound are also plotted. (a)~Rate $R_{\rm BMST}=0.49$ BMST-R $[2,1]$ codes with encoding memory $\Memory=8$ and coupling length $L=392$. The Cartesian product orders of the two BMST-R codes are $\CarBlocksBMST=750$ and $\CarBlocksBMST=1500$, respectively. (b)~A rate $R_{\rm BMST}=0.74$ BMST-SPC $[4,3]$ code with encoding memory $\Memory=4$, coupling length $L=296$, and Cartesian product order $\CarBlocksBMST=1200$.}
\label{Finite_BMST_RC_m8d8_DiffM}
\ifCLASSOPTIONonecolumn
\end{figure*}
\fi
\ifCLASSOPTIONtwocolumn
\end{figure}
\fi

\begin{example}[Finite-Length Performance]
Consider rate $R_{\rm BMST}=0.49$ BMST-R $[2,1]$ codes with $\Memory=8$ and $L=392$. The BER performance of BMST-R codes decoded with different decoding delays $d$ is shown in Fig.~\ref{Finite_BMST_RC_m8d8_DiffM}(a), where we observe that
\begin{enumerate}
  \item The BER performance of BMST-R codes decoded with different delays $d$ matches well with the corresponding window decoding thresholds in the high SNR region.
  \item The BER performance in the waterfall region improves as the decoding delay $d$ increases, but it does not improve much further beyond a certain decoding delay~(roughly $d=10$).
  \item The error floor improves as the decoding delay $d$ increases, and it matches well with the lower bound for BMST-R codes with $\Memory=8$ when $d$ increases up to a certain point~(roughly $d=16$).
\end{enumerate}
These results are consistent with the asymptotic threshold performance analysis shown in Fig.~\ref{Threshold_RC21_SPC43_DiffDelay}(a).

Similar behavior has also been observed for BMST-SPC code ensembles, as shown in Fig.~\ref{Finite_BMST_RC_m8d8_DiffM}(b), where the simulated decoding performance of a rate $R_{\rm BMST}=0.74$ BMST-SPC $[4,3]$ code constructed with $\Memory=4$, $L=296$, and $\CarBlocksBMST=1200$, and decoded with different decoding delays $d$ is depicted.
\end{example}

\subsection{Fixed $\Memory$ and $d$, Increasing $\CarBlocksBMST$}
\ifCLASSOPTIONonecolumn
\begin{figure*}[t]
\fi
\ifCLASSOPTIONtwocolumn
\begin{figure}[t]
\fi
\centering
\subfigure[~]{
\label{RC_m8d8}
\includegraphics[clip, width=\figwidth]{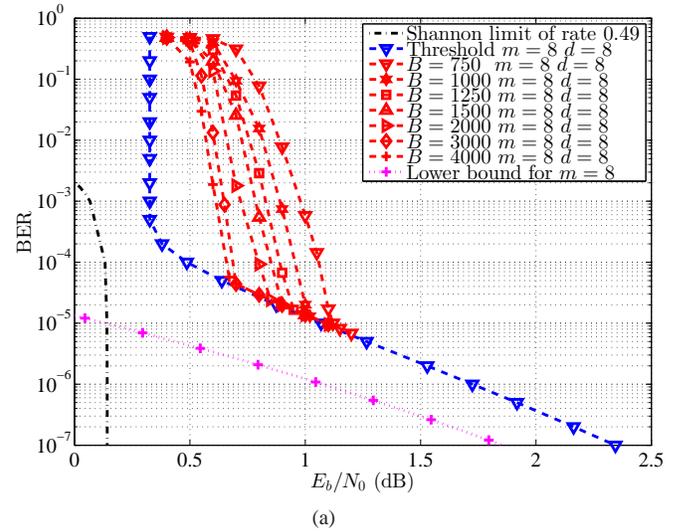}}
\subfigure[~]{
\label{RC_m8d16}
\includegraphics[clip, width=\figwidth]{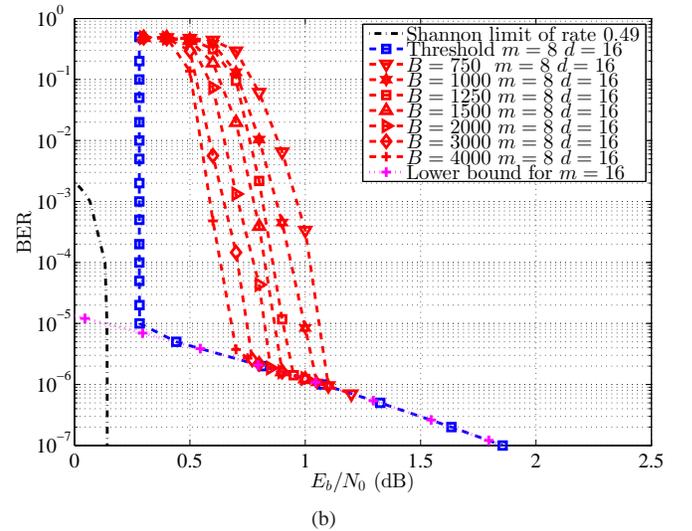}}
\caption{Simulated decoding performance of rate $R_{\rm BMST}=0.49$ BMST-R $[2,1]$ codes with different Cartesian product orders $B$. The encoding memory $\Memory=8$ and the coupling length $L=392$. The codes are decoded with (a)~decoding delay $d=8$, and (b)~$d=16$. The corresponding window decoding thresholds and the lower bound for BMST-R codes with $\Memory=8$ are also plotted.}
\label{Finite_BMST_RC_m8_DiffM}
\ifCLASSOPTIONonecolumn
\end{figure*}
\fi
\ifCLASSOPTIONtwocolumn
\end{figure}
\fi
\begin{example}[Finite-Length Performance]
Consider rate $R_{\rm BMST}=0.49$ BMST-R $[2,1]$ codes with $\Memory=8$ and $L=392$. The BER performance of BMST-R codes constructed with different Cartesian product orders $\CarBlocksBMST$ is shown in Fig.~\ref{Finite_BMST_RC_m8_DiffM}, where we observe that
\begin{enumerate}
  \item Similar to SC-LDPC codes, where increasing the lifting factor $\CarBlocks$ improves waterfall region performance, increasing the Cartesian product order $\CarBlocksBMST$ of BMST codes also improves waterfall region performance. As expected, this improvement saturates for sufficiently large $\CarBlocksBMST$. For example, the improvement at a BER of $10^{-5}$ from $B=1000$ to $B=2000$, both decoded with $d=16$, is about 0.17 dB, while the improvement decreases to about 0.06 dB from $B=3000$ to $B=4000$.

  \item The BER performance of BMST-R codes matches well with the corresponding window decoding thresholds in the error floor region.

  \item The error floors, which are solely determined by the encoding memory $\Memory$~(see Section~\ref{SecIII_Bound}), cannot be lowered by increasing $\CarBlocksBMST$.
\end{enumerate}
\end{example}

\textbf{Remark:} We found from simulations that, in the error floor region, the gap between finite-length performance and window decoding threshold ${({E_b}/{N_0})}^*$ is less than 0.02 dB. For example, the values of $E_b/N_0$ needed to achieve a BER of $10^{-5}$ for a BMST-R $[2,1]$ code with $\Memory=8$, very extremely large Cartesian product order (say, $\CarBlocksBMST=4000$), and decoding delay $d=8$ is 1.087 dB, while the calculated window decoding threshold for a preselected target BER of $10^{-5}$ of the BMST-R $[2,1]$ code ensemble with $\Memory=8$ and $d=8$ is ${({E_b}/{N_0})}^*=1.069$. This result again demonstrates that the finite-length performance is consistent with the asymptotic performance analysis.

\ifCLASSOPTIONonecolumn
\begin{figure*}[t]
\fi
\ifCLASSOPTIONtwocolumn
\begin{figure}[t]
\fi
\centering
\subfigure[~]{
\label{RC21_Diffm}
\includegraphics[clip, width=\figwidth]{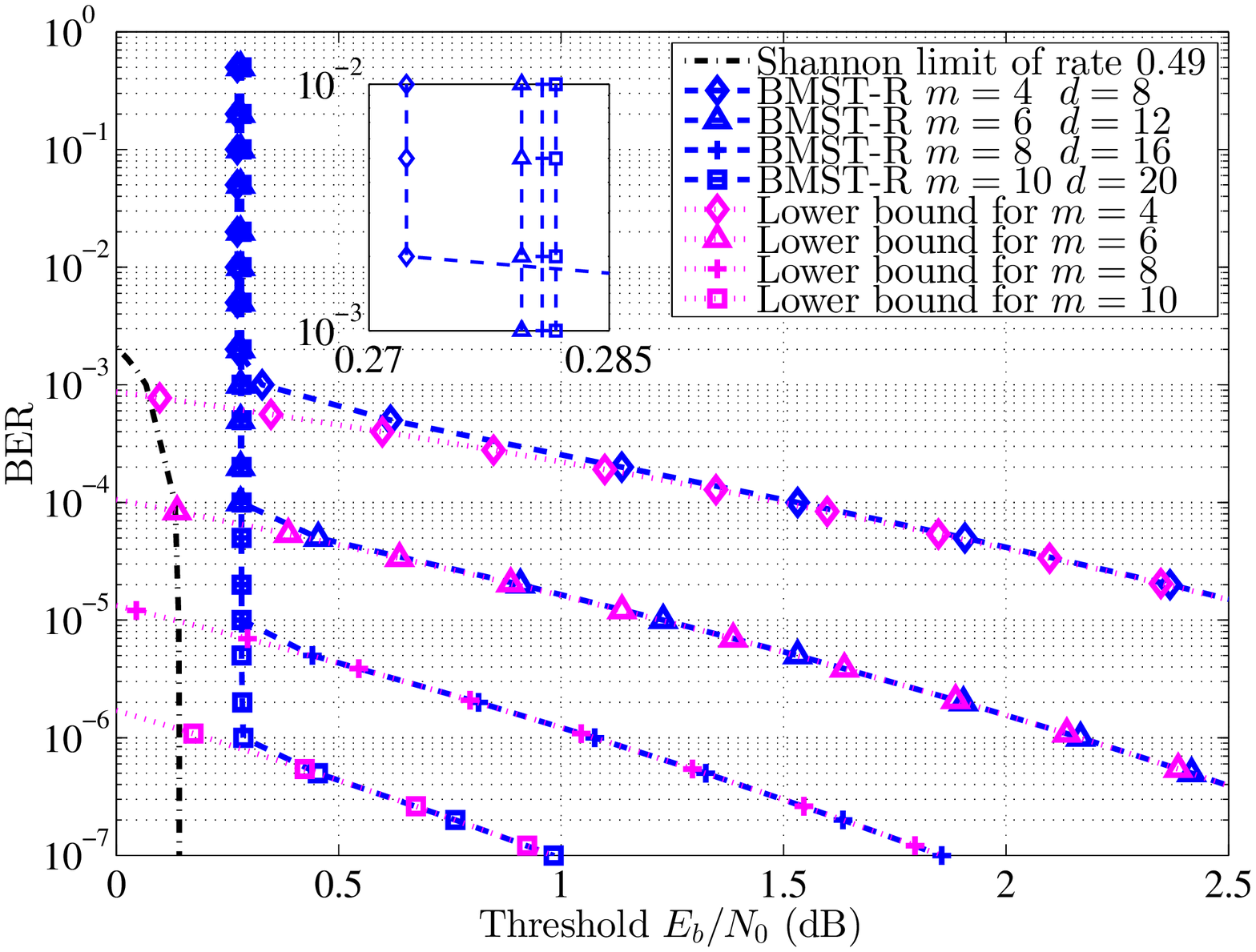}}
\subfigure[~]{
\label{SPC43_Diffm}
\includegraphics[clip, width=\figwidth]{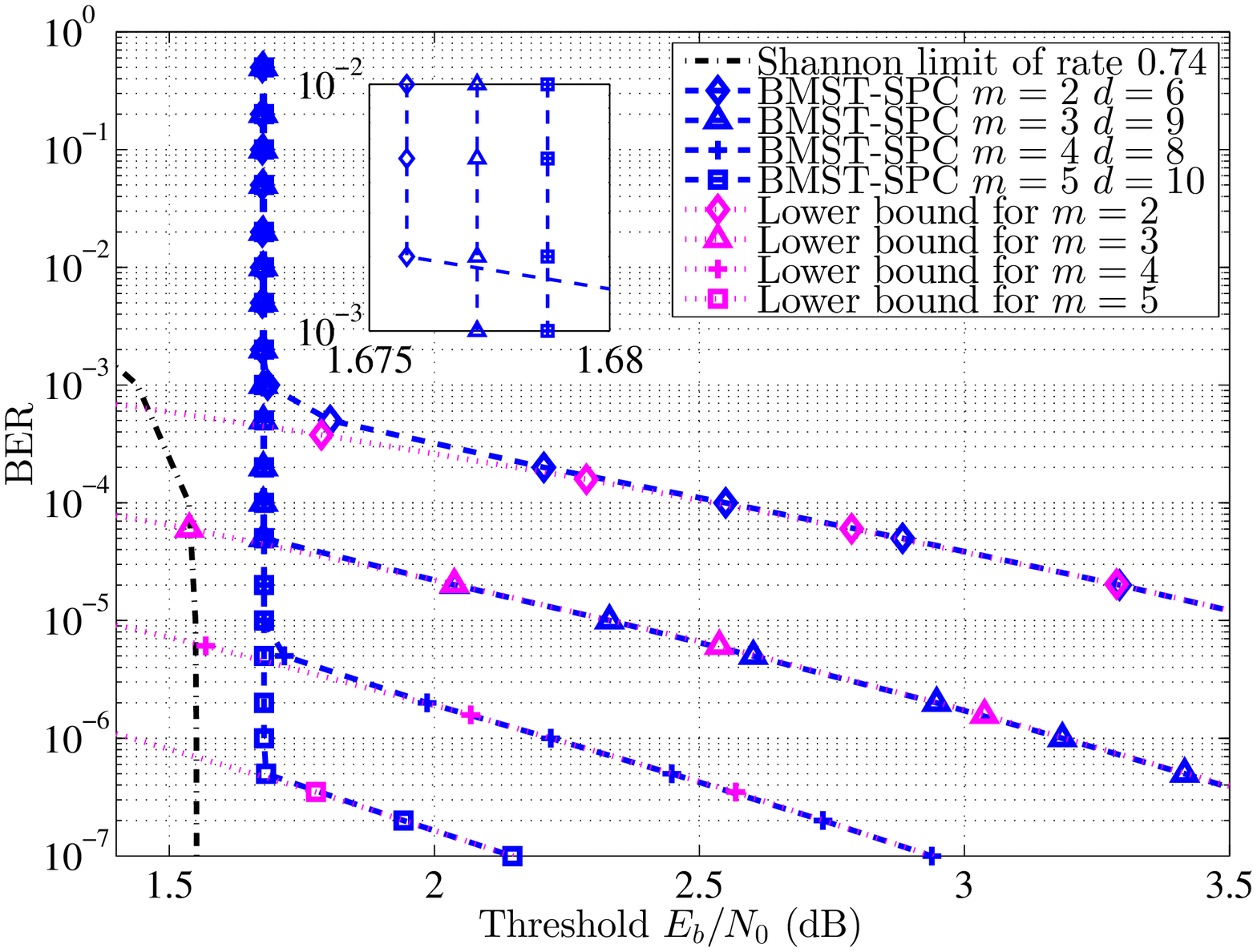}}
\caption{Window decoding thresholds in terms of ${({E_b}/{N_0})}^*$ (dB) with different target BERs for two BMST code ensemble families with different encoding memories $\Memory$. (a)~BMST-R $[2,1]$ code ensemble and (b)~BMST-SPC $[4,3]$ code ensemble.}
\label{Threshold_RC21_SPC43_Diffm}
\ifCLASSOPTIONonecolumn
\end{figure*}
\fi
\ifCLASSOPTIONtwocolumn
\end{figure}
\fi
\subsection{Fixed $\CarBlocksBMST$, Increasing $\Memory$~(and hence $d$)}
\begin{example}[Asymptotic Performance]
Consider a family of $R_{\rm BMST}=0.49$ BMST-R $[2,1]$ code ensembles with different encoding memories $\Memory$. The calculated window decoding thresholds in terms of the SNR ${E_b}/{N_0}$ versus the preselected target BERs together with the lower bounds are shown in Fig.~\ref{Threshold_RC21_SPC43_Diffm}(a), where we observe that
\begin{enumerate}
  \item For a high target BER~(roughly above $10^{-3}$), the threshold with a sufficiently large decoding delay degrades slightly as the encoding memory $\Memory$ increases, due to errors propagating to successive decoding windows.
  \item The error floor can be lowered by increasing the encoding memory $m$~(and hence the decoding delay $d$).
\end{enumerate}

Similar behavior has also been observed for BMST-SPC code ensembles, as shown in Fig.~\ref{Threshold_RC21_SPC43_Diffm}(b), where the thresholds of a family of rate $R_{\rm BMST}=0.74$ BMST-SPC $[4,3]$ code ensembles are depicted.
\end{example}

\begin{figure}[t]
  \centering
  \includegraphics[width=\figwidth]{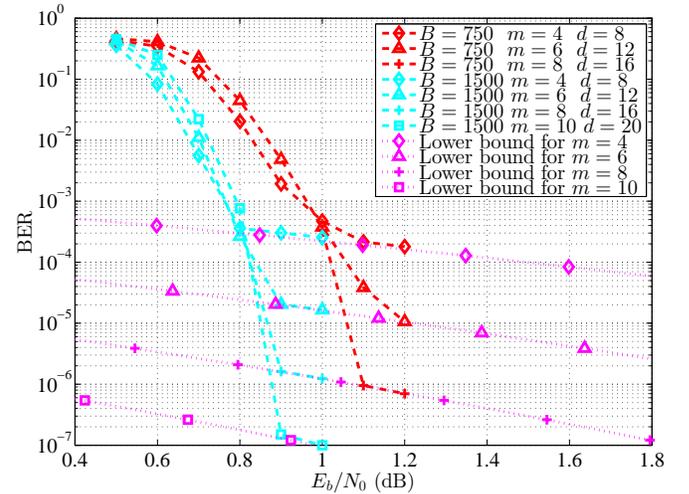}
  \caption{Simulated decoding performance of rate $R_{\rm BMST}=0.49$ BMST-R $[2,1]$ codes constructed with different encoding memories $\Memory$ and decoded with decoding delay $d=2\Memory$. The Cartesian product orders of the involved BMST-R codes are $\CarBlocksBMST=750$ and $\CarBlocksBMST=1500$. The corresponding lower bound for BMST-R codes is also plotted.}
  \label{Finite_BMST_RC_SameM_Diffm}
\end{figure}
\begin{example}[Finite-Length Performance]
Consider rate $R_{\rm BMST}=0.49$ BMST-R $[2,1]$ codes constructed with encoding memories $\Memory=4$, $6$, $8$, and $10$, and Cartesian product orders $\CarBlocksBMST=750$ and $\CarBlocksBMST=1500$. The simulated BER performance with sufficiently large decoding delay is shown in Fig.~\ref{Finite_BMST_RC_SameM_Diffm}, where we observe that
\begin{enumerate}
  \item The BER performance in the waterfall region degrades slightly as the encoding memory $\Memory$ increases, due to errors propagating to successive decoding windows.
  \item The error floor of the BER curves is lowered by increasing the encoding memory $\Memory$~(and hence the decoding delay $d$).
\end{enumerate}
These results are consistent with the asymptotic performance analysis shown in Fig.~\ref{Threshold_RC21_SPC43_Diffm}(a).
\end{example}

\section{Performance and Complexity Comparison of SC-LDPC Codes and BMST Codes}\label{SecVI}
In addition to decoding performance, the latency introduced by employing channel coding is a crucial factor in the design of a practical communication system. For example, minimizing latency is very important in applications such as personal wireless communication and real-time audio and video. In this section, we first compare the performance of BMST codes and SC-LDPC codes when the two decoding latencies are equal. Then a computational complexity comparison is presented.

We restrict consideration to $(3,6)$-regular SC-LDPC codes with coupling width $\Memory=1$, where two component submatrices $\mathbf{B}_{0}=[2~1]$ and $\mathbf{B}_{1}=[1~2]$ are used, due to their superior thresholds and finite-length performance with window decoding when the decoding delay is relatively small~(see, e.g.,~\cite{Wei14_IT,Huang14}). For the BMST codes, we consider BMST-R [2,1] codes with encoding memory $\Memory=8$, due to their near-capacity performance in the waterfall region and relatively low error floor~(see Section~\ref{SecV}). In the simulations, the iterative sliding window decoding algorithm for SC-LDPC codes uses the uniform parallel~(flooding) updating schedule with a maximum iteration number of 100, while for the BMST codes, window decoding is performed using the layer-by-layer updating schedule with a maximum iteration number of 18. The entropy stopping criterion~\cite{Ma04,Ma15} is employed for both window decoding algorithms with a preselected threshold of $10^{-6}$.

The decoding latency of the sliding window decoder, in terms of bits, is given by~\cite{Huang14}
\vspace{\vspaceh}
\begin{equation}\label{SC_LDPC_Latency}
    T_{\rm SC}=2M(d_{\rm SC}+1)
\end{equation}
for the $(3,6)$-regular SC-LDPC codes, and
\vspace{\vspaceh}
\begin{equation}\label{BMST_RC_latency}
    T_{\rm BMST}=2B(d_{\rm BMST}+1)
\end{equation}
for the BMST-R $[2,1]$ codes, where $d_{\rm SC}$ and $d_{\rm BMST}$ are the decoding delays of the SC-LDPC codes and BMST codes, respectively. When the parameters $\CarBlocks$, $\CarBlocksBMST$, $d_{\rm SC}$, and $d_{\rm BMST}$ satisfy $\CarBlocksBMST=\CarBlocks(d_{\rm SC}+1)/(d_{\rm BMST}+1)$, the decoding latency of BMST-R $[2,1]$ codes is the same as that of $(3,6)$-regular SC-LDPC codes. In our simulations, we consider decoding delay $d_{\rm SC}=5$~(i.e., window size $W=d_{\rm SC}+1=6$), which is a good choice for the SC-LDPC codes to achieve optimum performance when the decoding latency is fixed~\cite{Huang14}.
\subsection{Performance Comparison}

\begin{figure}[t]
  \centering
  \includegraphics[width=\figwidth]{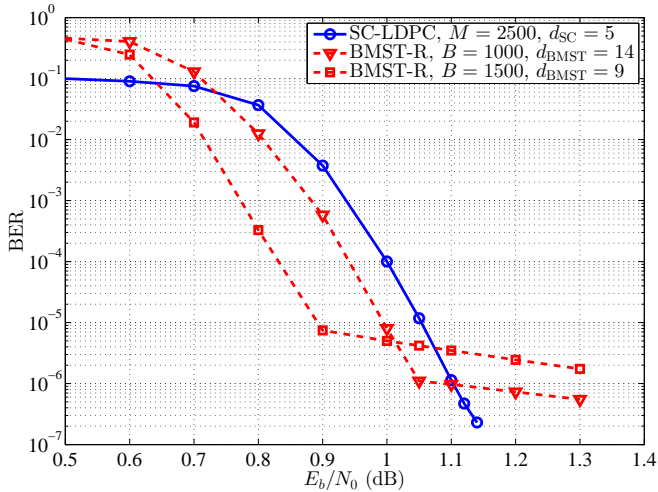}
  \caption{Simulated decoding performance of BMST-R $[2,1]$ codes with encoding memory $\Memory=8$ compared to $(3,6)$-regular SC-LDPC code with coupling width $\Memory=1$. The values of $B$ and $d_{\rm BMST}$ for the BMST-R codes are chosen in such a way that the decoding latencies of all the codes are the same.}
  \label{Finite_BMST_SCLDPC_SameLatency}
\end{figure}
In Fig.~\ref{Finite_BMST_SCLDPC_SameLatency}, BMST-R $[2,1]$ codes are compared to $(3,6)$-regular SC-LDPC codes, where the values of the Cartesian product order $B$ and decoding delay $d_{\rm BMST}$ for the BMST-R codes are chosen such that the two decoding latencies $T_{\rm BMST}$ and $T_{\rm SC}$ are the same. We see that the BMST-R codes outperform the SC-LDPC code in the waterfall region but have a higher error floor. From Fig.~\ref{Finite_BMST_SCLDPC_SameLatency}, we also see that, in the waterfall region, the BMST-R code constructed with a larger Cartesian product order $B$ and decoded with a smaller decoding delay $d_{\rm BMST}=9$ outperforms the BMST-R code constructed with a smaller $B$ and decoded with a larger decoding delay $d_{\rm BMST}=14$ but has a higher error floor~(both have the same decoding latency). In other words, selecting a smaller $d_{\rm BMST}$, which is typically detrimental to decoder performance, is compensated for by allowing a larger $B$, which improves code performance. For example, at a BER of $10^{-5}$, the BMST-R code with $B=1000$ and decoded with decoding delay $d_{\rm BMST}=14$ gains $0.05$ dB compared to the equal latency SC-LDPC code with $M=2500$, while the gain increases to $0.15$ dB by using the BMST-R code with $B=1500$ and $d_{\rm BMST}=9$.

\begin{figure}[t]
        \center
        \includegraphics[clip, width=\figwidth]{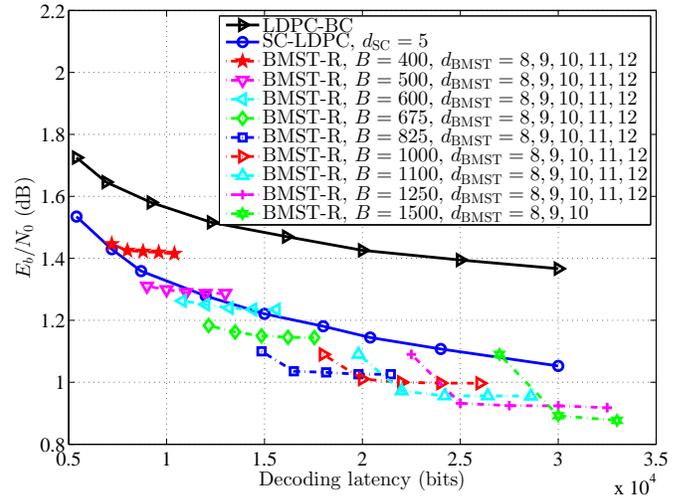}
        \caption{Required $E_b/N_0$ to achieve a BER of $10^{-5}$ for $(3,6)$-regular LDPC-BCs, $(3,6)$-regular SC-LDPC codes, and BMST-R $[2,1]$ codes as a function of decoding latency.}
        \label{FunctionLatency}
\end{figure}
The $E_b/N_0$ required to achieve a BER of $10^{-5}$ for equal latency $(3,6)$-regular LDPC-BCs, $(3,6)$-regular SC-LDPC codes, and BMST-R $[2,1]$ codes as a function of decoding latency is shown in Fig.~\ref{FunctionLatency}, where we observe that both the BMST-R codes and the SC-LDPC codes perform significantly better than the LDPC-BCs. Also, the performance of the BMST-R codes~(with fixed Cartesian product order $B$) improves as the decoding delay $d_{\rm BMST}$~(and hence the latency) increases, but it does not improve much further beyond a certain decoding delay~(roughly $d_{\rm BMST}=10$). (Note again that increasing the decoding delay $d_{\rm BMST}$ improves decoder performance and increasing the Cartesian product order $B$ improves code performance.) However, under an equal decoding latency assumption, increasing the decoding delay $d_{\rm BMST}$ or the Cartesian product order $B$ does not always lower the $E_b/N_0$ required to achieve a BER of $10^{-5}$. For example, when the decoding latency is $14850$ bits, the performance of the BMST-R code with $B=825$ and decoded with $d_{\rm BMST}=8$ is better than that of the BMST-R code with $B=675$ and decoded with $d_{\rm BMST}=10$. However, if we increase the latency to 19800 bits, the code with the Cartesian product order $B=825$ and decoded with a larger $d_{\rm BMST}=11$ still outperforms the code with $B=1100$ and decoded with a smaller $d_{\rm BMST}=8$. This raises the interesting question of how to choose $B$ and $d_{\rm BMST}$ in order to achieve the best performance when the decoding latency of the sliding window decoder for BMST-R codes is fixed.

\begin{figure}[t]
        \center
        \includegraphics[clip, width=\figwidth]{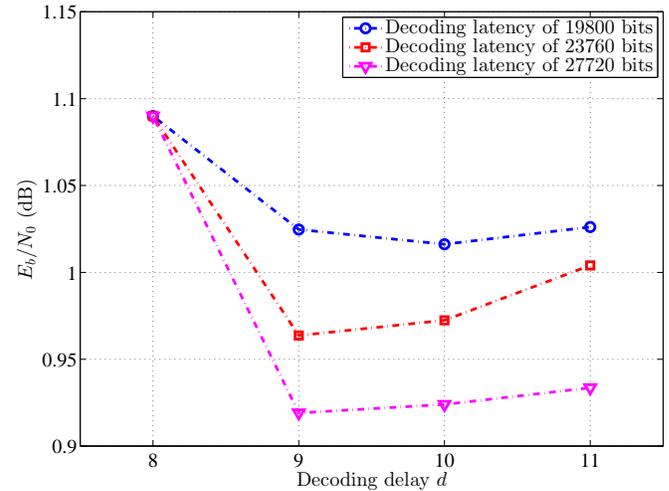}
        \caption{Required $E_b/N_0$ to achieve a BER of $10^{-5}$ for BMST-R $[2,1]$ codes with different decoding delays $d_{\rm BMST}$ and decoding latencies of 19800, 23760, and 27720 bits.}
        \label{Finite_BMST_SameLatency_DiffDelay}
\end{figure}

We also see from Fig.~\ref{FunctionLatency} that, for a fixed decoding latency roughly less than 15000 bits, to achieve a BER of $10^{-5}$, $d_{\rm BMST}=8$ is a good choice for optimum performance. This is due to the fact that the interleavers, which break short cycles in the normal graph of BMST codes, especially when the interleavers of size $n=\CarN\CarBlocksBMST$ are generated randomly, play a crucial role in iterative decoding~\cite{Ma15}. That is, the larger the Cartesian product order $B$ is, the better the performance of BMST codes becomes. However, the value of $E_b/N_0$ required to achieve a BER of $10^{-5}$ for BMST-R $[2,1]$ codes decoded with a fixed decoding delay $d_{\rm BMST}$ is bounded below by its corresponding window decoding threshold~(see Section~\ref{SecV}-B).

Fig.~\ref{Finite_BMST_SameLatency_DiffDelay} shows the $E_b/N_0$ values required for BMST-R [2,1] codes to achieve a BER of $10^{-5}$ with different decoding delays $d_{\rm BMST}$ and larger decoding latencies of 19800, 23760, and 27720 bits. Here we see that the required values of $E_b/N_0$ for the BMST-R $[2,1]$ codes with $d_{\rm BMST}=8$ are the same and approach the corresponding window decoding threshold~(as remarked in Section~\ref{SecV}-B). In this case, however, we also observe that the required values of $E_b/N_0$ continue to decreases until roughly $d_{\rm BMST}=9\thicksim10$, and then they increase gradually as the decoding delay $d_{\rm BMST}$ increases further. This increase results from the fact that the improved decoder performance obtained by increasing $d_{\rm BMST}$ is not compensating for the decrease in code performance as a result of the smaller Cartesian product order $B$. Thus, for larger decoding latencies~(up to 35000 bits), $d_{\rm BMST}=9$ is a good choice for optimum performance.

\subsection{Complexity Comparison}
As shown in~\cite{Ma15}, we can measure the computational complexity of BMST codes by the total number of operations. Consider a BMST-R $[\CarN,1]$ code or a BMST-SPC $[\CarN,\CarN-1]$ code with Cartesian product order $B$ and decoding delay $d_{\rm BMST}$. Let $Opt(\mathcal{A})$ denote the number of operations at a generic node $\mathcal{A}$. Each decoding layer has $\CarN B$ parallel nodes \fbox{$=$}, $\CarN B$ parallel nodes \fbox{$+$}, and a node of type \fbox{G}. The computational complexity for each node \fbox{$=$}, each node \fbox{$+$}, and each node \fbox{G} is $\mathcal{O}(m+2)$, $\mathcal{O}(m+1)$, and $\mathcal{O}(\CarN B)$, respectively. Thus, the total number of operations for each decoding layer update is given by
\vspace{\vspaceh}
\begin{equation}\label{BMST_ComplexityPerLayer}
\begin{array}{l}
    \CarN B \cdot Opt\left(\fbox{=}\right) + \CarN B \cdot Opt\left(\fbox{+}\right) + Opt\left(\fbox{G}\right) \\
    = \CarN B (m+2) + \CarN B (m+1) + \CarN B = \CarN B (2m+4).
\end{array}
\end{equation}
Let $I_{\rm BMST}$ denote the average number of iterations required to decode a target layer for BMST codes. Since each iteration requires both a forward recursion~($d_{\rm BMST}$ layer-updates) and a backward recursion~($d_{\rm BMST}$ layer-updates), the total~(average) computational complexity per window is given by
\vspace{\vspaceh}
\begin{equation}\label{BMST_ComplexityPerWindow}
\begin{array}{l}
\mathcal{O}( \CarN B (2m+4)\times 2d_{\rm BMST} )I_{\rm BMST}\\= \mathcal{O}( \CarN B (4m+8) d_{\rm BMST} )I_{\rm BMST}.
\end{array}
\end{equation}
Note that the number of decoded~(target) bits for the window decoder at each time instant is $\CarN B$, and thus the computational complexity per decoded bit for a BMST code is
\vspace{\vspaceh}
\begin{equation}\label{BMST_ComplexityPerBit}
\begin{array}{l}
   \mathcal{O}( \CarN B (4m+8) d_{\rm BMST} )I_{\rm BMST}/(\CarN B)\\=\mathcal{O}\left( (4m+8) d_{\rm BMST} I_{\rm BMST} \right).
\end{array}
\end{equation}

Now consider a $(3,6)$-regular SC-LDPC code with lifting factor $M$ and decoding delay $d_{\rm SC}$~(the corresponding decoding window size $W=d_{\rm SC}+1$). Let $I_{\rm SC}$ denote the average number of iterations required to decode a target layer for SC-LDPC codes. Note that the numbers of operations at a variable node and a check node of $(3,6)$-regular SC-LDPC codes are 3 and 6, respectively. The average computational complexity~(also measured by the total number of operations) per window is then given by
\vspace{\vspaceh}
\begin{equation}\label{SC_Complexity}
    \mathcal{O}\left( 3 T_{\rm SC} + 6 T_{\rm SC}/2 \right) I_{\rm SC}=\mathcal{O}\left( 6T_{\rm SC}I_{\rm SC} \right),
\end{equation}
where $T_{\rm SC}$ is the decoding latency. Note that the number of decoded~(target) bits for the window decoder at each time instant is $T_{\rm SC}/(d_{\rm SC}+1)$, and thus the computational complexity per decoded bit for a $(3,6)$-regular SC-LDPC code is
\vspace{\vspaceh}
\begin{equation}\label{SC_ComplexityPerBit}
    \frac{\mathcal{O}\left( 6T_{\rm SC} \right) I_{\rm SC}}{T_{\rm SC}/(d_{\rm SC}+1)}=\mathcal{O}\left( 6(d_{\rm SC}+1) I_{\rm SC} \right).
\end{equation}

\begin{table*}[t]
\caption{Computational complexity per decoded bit of a $(3,6)$-regular SC-LDPC code and BMST-R $[2,1]$ codes that achieve a BER of $10^{-5}$ with decoding latency of 30000 bits}\label{Table2}
  \centering
  \begin{tabular}{|c||c|c|c|c|c|}
  \hline
  Codes   &~~$M\backslash B$~~ &~~~$m$~~~ &$d_{\rm SC}\backslash d_{\rm BMST}$ &$I_{\rm SC}\backslash I_{\rm BMST}$ &~Complexity\\ \cline{1-6}
  SC-LDPC &2500    &1   &5   &9.65                    &347.4 \\ \hline
  BMST    &1000    &8   &14  &2.03                    &1136.8 \\ \hline
  BMST    &1500    &8   &9   &3.20                    &1152.0 \\ \hline
\end{tabular}
\end{table*}
Table~\ref{Table2} shows the average computational complexity per decoded bit of the $(3,6)$-regular SC-LDPC code and the BMST-R $[2,1]$ codes used in Fig.~\ref{Finite_BMST_SCLDPC_SameLatency} that achieve a BER of $10^{-5}$ with a decoding latency of 30000 bits. The simulation parameters $M$, $B$, $m$, $d_{\rm BMST}$, $d_{\rm SC}$, $I_{\rm BMST}$ and $I_{\rm SC}$ are also included. We observe that, though the average number of iterations $I_{\rm BMST}$ for the BMST code is significantly less than $I_{\rm SC}$ for SC-LDPC code, the computational complexity per decoded bit for the BMST codes is higher than for the SC-LDPC code. However, the BMST codes outperform the SC-LDPC code in the waterfall region (see Fig.~\ref{Finite_BMST_SCLDPC_SameLatency} in Section~\ref{SecVI}-B). This means that BMST-R $[2,1]$ codes, compared to $(3,6)$-regular SC-LDPC codes, obtain performance gains at a cost of higher computational complexity.

\section{Conclusions}\label{sec:Conclusion}
In this paper, we described BMST codes using both an algebraic description and a graphical representation for the purpose of showing that BMST codes can be viewed as a class of SC codes. Then, based on a modified EXIT chart analysis and finite-length computer simulations, we investigated the impact of several parameters~(coupling width, Cartesian product order, and decoding delay) on the performance of BMST codes. We then examined the relationship between the Cartesian product order, the decoding delay, and the decoding performance of BMST codes for fixed decoding latency in comparison to SC-LDPC codes, and a comparison of computational complexity was also presented. It was observed that, under the equal decoding latency constraint, BMST codes using the repetition $[2,1]$ code~(BMST-R $[2,1]$ code) as the basic code can outperform $(3,6)$-regular SC-LDPC codes in the waterfall region but have a higher error floor and a larger decoding complexity. An interesting future research topic to complement the work reported here is to embed a partial superposition strategy into the code design to further improve the performance of the original BMST codes for a given decoding latency.

\section*{Acknowledgment}
The authors would like to thank Prof. Daniel J. Costello, Jr. for his helpful comments, polishing this paper,  and invaluable contributions as a co-author of the conference version of this paper~\cite{Huang15_ISIT}. They would also like to thank Mr. Chulong Liang from Sun Yat-sen University for helpful discussions.


\ifCLASSOPTIONcaptionsoff
  \newpage
\fi

\end{document}